\documentclass[12pt,preprint]{aastex}
\usepackage{epsfig}
\usepackage{psfig}
\usepackage{graphicx}
\def\aa{{A\&A}}

\def\aj{{AJ}}
\def\al{$\alpha$}

\def\amin{$^\prime$}
\def\annrev{{ARA\&A}}
\def\apj{{ApJ}}
\def\apjs{{ApJS}}
\def\asec{$^{\prime\prime}$}

\def\cc{cm$^{-3}$}
\def\deg{$^{\circ}$}

\def\cc{cm$^{-3}$}

\def\h{\hskip -4 mm}

\def\lamb{$\lambda$}
\def\lax{{$\mathrel{\hbox{\rlap{\hbox{\lower4pt\hbox{$\sim$}}}\hbox{$<$}}}$}}
\def\gax{{$\mathrel{\hbox{\rlap{\hbox{\lower4pt\hbox{$\sim$}}}\hbox{$>$}}}$}}
\def\simlt{\lower.5ex\hbox{$\; \buildrel < \over \sim \;$}}
\def\simgt{\lower.5ex\hbox{$\; \buildrel > \over \sim \;$}}
\def\lum{erg s$^{-1}$}

\def\mnras{{MNRAS}}

\def\pasp{{PASP}}

\def\percm2{cm$^{-2}$}
\def\perhz{Hz$^{-1}$}
\def\peryr{yr$^{-1}$}

\def\solum{$L_\odot$}
\def\solmass{$M_\odot$}
\def\oii{[\ion{O}{2}]}

\def\hii{\ion{H}{2}}
\def\oiii{[\ion{O}{3}]}

\def\nd{\nodata}

\slugcomment{Submitted to {\it The Astronomical Journal}}
\shorttitle{Radio Properties of Type 2 Quasars}
\shortauthors{Lal \& Ho}

\begin{document}

\title{The Radio Properties of Type 2 Quasars}

\author{Dharam Vir Lal$^{1,2}$ and Luis C. Ho$^{3}$}
\affil{$^1$ Institute of Astronomy \& Astrophysics, Academia Sinica, PO Box 23-141, Taipei 10617, Taiwan}
\affil{$^2$ Harvard-Smithsonian Center for Astrophysics, 60 Garden Street, MS-67, Cambridge, MA 02138, USA}
\affil{$^3$ The Observatories of the Carnegie Institution for Science, 813 Santa Barbara Street, Pasadena, CA 91101, USA}

\begin{abstract}
This paper presents the first high-resolution and high-sensitivity study of the 
radio properties of optically selected type~2 quasars.  We used the Very Large 
Array at 8.4 GHz to observe 59 sources drawn from the Sloan Digital Sky Survey 
sample of Zakamska et~al. (2003).  The detection rate of our survey is 59\% 
(35/59), comparable to the detection rate in FIRST at 1.4 GHz.  Ongoing star 
formation, although present, contributes negligible radio emission at the 
current sensitivity limit.  Comparing the radio powers with the \oiii\ 
\lamb5007 luminosities, we find that roughly 15\%$\pm$5\% of the sample can be 
considered radio-loud.  Intriguingly, the vast majority of 
the detected sources in our sample fall in a region intermediate between those 
traditionally occupied by radio-loud and radio-quiet quasars.  Moreover, most 
of these ``radio-intermediate'' sources tend to have flat or inverted radio 
spectra, which we speculate may be caused by free-free absorption by ionized 
gas in the narrow-line region.  The incidence of flat-spectrum sources in 
type~2 quasars appears to be much higher than in type~1 quasars, in
apparent violation of the simple orientation-based unified model for active 
galaxies.
\end{abstract}


\section{Introduction}
\label{sec:intro}

The collective properties of active galactic nuclei (AGNs) can be understood 
within the framework of unification models in which the observed properties
of active galaxies are governed primarily by orientation, intrinsic luminosity, 
and Eddington ratio (Antonucci 1993; Urry \& Padovani 1995; Ho 2008).  Many of 
the apparent differences between type~1 (broad-line) and type~2 (narrow-line) 
AGNs can be attributed to our line-of-sight having different orientations with 
respect to the disk.  In conventional unified models of active galaxies, the 
accretion disk and broad-line region are surrounded by a dusty molecular torus. 
In type~2 AGNs, where the view being edge-on, the torus blocks a direct view of 
the continuum and broad-line region, which can only be detected through light 
scattered into the line-of-sight by material lying directly above the torus 
opening; by contrast, in type~1 AGNs, the view being pole-in, the continuum and 
broad-line region can be viewed directly.  The basic predictions of this 
geometric unification scheme have enjoyed much success in the context of 
Seyfert galaxies (e.g., Antonucci 1993; Lal et al. 2004).  A natural question 
is whether this simple picture extends to AGNs of higher luminosity: are there
type~2 quasars?  

As an alternative to the classical orientation-based unification scheme, 
wherein the obscuration arises solely from a nuclear-scale dusty torus, the 
obscuration in type~2 quasars may largely come instead from the large-scale 
distribution of dust in the host, which is itself tied to the evolutionary 
state of the galaxy.  For example, the hierarchical buildup of galaxies through
major mergers of gas-rich progenitors naturally produces remnants with copious 
stockpiles of centrally concentrated gas and dust, which not only fuels 
a major starburst but also often accretion onto a highly obscured AGN 
(e.g., Sanders \& Mirabel 1996; Genzel et al. 1998).  If sufficiently 
powerful, the buried AGN in these ultraluminous infrared systems would 
qualify as type~2 quasars.  Over time, as the gas and dust get depleted
through consumption and expulsion from energy feedback, the AGN transitions 
from an obscured to an unobscured source (e.g., Sanders et al. 1988; Hopkins 
et al. 2006).  In this picture, type~2 quasars are the evolutionary 
precursors of, and {\it intrinsically}\ different from, type~1 quasars.

A large body of work over the last several years has established beyond doubt 
that type~2 quasars truly do exist.  Individual case studies (e.g., Dawson et 
al. 2001; Norman et al. 2002; Stern et al. 2002; Derry et al. 2003; Jarvis et 
al. 2005) have given way to large systematic searches for obscured quasars from 
hard X-ray (e.g., Ueda et al. 2003; Sazonov et al. 2007), infrared (e.g., Lacy 
et al.  2004; Polletta et al. 2008), and optical (Zakamska et al. 2003; Dong et 
al. 2005; Reyes et al. 2008) surveys.  The Sloan Digital Sky Survey (SDSS; York 
et al. 2000) makes it possible to find substantial numbers of optically 
selected type~2 quasar candidates.  The first results from Zakamska et~al. (2003)
produced a large sample of 291 objects at redshift 0.3 $<$ $z$ $<$ 0.83;
the sample includes Seyfert 2 galaxies as well as objects luminous enough
to qualify as type~2 quasar candidates.  The latest update from Reyes et~al. (2008)
has increased the list of candidates to nearly 900 sources.  The 
overall concensus seems to be that type~2 quasars constitute an important 
component of the general AGN population.  In a related development, radio 
and infrared selection have revealed increasingly large numbers of red 
quasars (e.g., Webster et al. 1995; Gregg et al. 
2002; Richards et al. 2003; White et al. 2003), whose nonstandard colors 
in most cases can be attributed to increased dust reddening (e.g., Glikman et 
al. 2007), perhaps linked to galaxy mergers (Urrutia et al. 2008).
The prevalence of high-luminosity 
obscured or dusty AGNs has important implications for a wide range of broader 
astrophysical issues, ranging from testing the classical unification model of 
AGNs to understanding the origin of the cosmic X-ray background (e.g., Madau 
et al. 1994; Comastri et al. 1995), the accretion history of supermassive 
black holes (e.g., Mart\'\i nez-Sansigre \& Taylor 2009), and the coevolution 
of black holes and galaxies (e.g., Greene et al. 2009).

To better understand the physical characteristics of type~2 quasars and their 
relation to other classes of AGNs, it is essential to define the multiwavelength
properties of these objects.  To date, follow-up studies have concentrated 
mostly on the initial SDSS sample of Zakamska et al. (2003), investigating 
their optical spectropolarimetric (Zakamska et al. 2005), emission-line 
(Villar-Mart\'\i n et al. 2008), X-ray (Vignali et al. 2004, 2006; Ptak et al. 
2006), infrared (Zakamska et al. 2008), and host galaxy (Zakamska et al. 2006; 
Greene et al. 2009; Liu et al. 2009) properties.  Zakamska et al. (2004) 
assembled sky-survey data at radio, infrared, optical, and soft X-ray 
wavelengths, concluding that the broadband properties of the sample support the 
thesis that the narrow-line SDSS sources are obscured (type~2) quasars.

This contribution presents new 8.4 GHz Very Large Array (VLA) observations 
of a subset of the SDSS type~2 quasars.  Notwithstanding the work of Zakamska 
et al. (2004), relatively little is actually known about the detailed radio 
properties of this new population of AGNs.  Using radio data primarily from the 
1.4 GHz FIRST survey (Becker et al. 1995), which has a root-mean-square (rms) 
sensitivity of $\sim$0.15 mJy, Zakamska et al. (2004) reported that less than 
half of the sample sources have radio counterparts.  At the resolution of FIRST 
($\sim$5\asec), the vast majority of the radio detections have a simple compact 
morphology.  Very limited information was given for the radio spectral 
shape, being limited to low-resolution data for the few brightest sources 
with extended morphologies.  

Our new 8.4 GHz observations represent a nearly ten-fold increase in 
sensitivity (rms $\approx 0.02$ mJy) and an improvement in resolution by 
more than a factor of 6 (beam $\sim$0\farcs8).  These improvements allow 
us to better delineate the basic radio properties of these sources, 
including their distribution of radio powers, source structure, and, 
in combination with the extant 1.4 GHz observations, spectral indices.

Distance-dependent quantities are calculated assuming $H_0$ = 70 
km~s$^{-1}$~Mpc$^{-1}$, $\Omega_m$ = 0.3, and $\Omega_\Lambda$ = 0.7.
Throughout the paper we define the spectral index $\alpha$ in the sense that 
$S_\nu \propto \nu^\alpha$, where $S_\nu$ is the flux density and $\nu$ is the 
frequency.

\section{Data}

\subsection{Sample} \label{sample}

Our targets are drawn from the type~2 quasar sample of Zakamska et~al.
(2003). This is the first comprehensive sample of such sources, for which we
understand the optical properties in great detail.  The SDSS sources have a
wide range of AGN power, ranging from luminosities in the regime of Seyfert 
galaxies to quasars.  Most of the sources qualify as quasars according to the 
\oiii\ $\lambda$5007 luminosity criterion of Zakamska et al. (2003).\footnote{
The traditional (albeit physically arbitrary) luminosity 
threshold for a quasar is $M_B({\rm Vega}) = -23$ mag for a cosmology of 
$H_0$ = 50 km~s$^{-1}$~Mpc$^{-1}$, $\Omega_m$ = 1, and $\Omega_\Lambda$ = 0 
(Schmidt \& Green 1983), which, in the modern cosmology adopted here and for 
$z = 0.3$, is equivalent to $M_B({\rm Vega}) = -22.5$ mag, or $M_B({\rm AB}) = 
-22.663$ mag. To relate this broadband absolute magnitude to an equivalent 
[O~{\tiny III}] luminosity, N. L.  Zakamska (private communications) recommends 
that we convert the continuum flux density in the $B$ band to 2500 \AA\ using 
the composite quasar slope of Vanden Berk et al. (2001), $f_\nu \propto 
\nu^{-0.44}$, and then use the empirical $L_{\rm [O~{\tiny III}]}-M_{2500}$ 
relation of Reyes et al. (2008; Equation 12).  The resulting line luminosity 
threshold for a quasar is then $L_{\rm [O~{\tiny III}]} = 7.8\times10^7$ \solum\ 
($3\times 10^{41}$ \lum).} Although the \oiii\ $\lambda$5007 emission might be 
mildly anisotropic (Hes et al. 1996; but see Kuraszkiewicz et al. 2000), it is 
widely used as a measure of intrinsic AGN power for both type~1 and type~2 
sources (e.g., Lal et~al. 2004; Heckman et al. 2005).  

Given the time allotted to our VLA program, we have decided to pick a subset of 
the SDSS sample in a restricted window in \oiii\ luminosity in order to 
explore its full range of radio properties.  Our final choice focuses on the 
59 sources, $\sim$20\% of the 291 sources in Zakamska et al. (2003), in the
luminosity range $L_{\rm [O~\sc III]} = 10^{41.8}-10^{42.1}$ \lum, which
coincides with the peak of the \oiii\ luminosity distribution and satisfies the
formal luminosity definition of type~2 quasars (Figure 1).  The sources are
distributed in the redshift range 0.3 \lax\ $z$ \lax\ 0.8, with an average
value of $z \approx 0.45$.

\subsection{Observations} \label{observation}

We were granted 24 hours of observing time for our program (program code
AV288).  The observations were carried out in snapshot mode in a single
observing run on 2006 July 24--25, with the VLA in B configuration, with two 50
MHz intermediate-frequency (IF) channels at a mean frequency of 8.4351 GHz (see
Table~1).  This provided a typical resolution of $\sim$
$0\hbox{$.\!\!^{\prime\prime}$}8$.  The weather conditions were excellent
during the observing session.  Since we are mainly concerned with the emission
from the source, which is placed at the phase-tracking center, bandwidth
smearing (Bridle \& Schwab 1999) is not a problem at 8.4 GHz despite the
somewhat large bandwidth.  The excellent sensitivity of the 8.4 GHz receivers
of the VLA routinely permits high signal-to-noise ratio maps to be made in
relatively short integration times (snapshot mode).  Each scan yielding a total
exposure time of $\sim$ 15$-$18 minutes for target source was interleaved with
2--3 minute scans on a suitable secondary VLA flux calibrator.  Due to the wide
range of right ascensions in the sample, some of the sources could only be 
observed when they were close to rising or setting, and consequently the 
synthesized beams were highly distorted.  More than half the phase calibrators 
used have position
accuracies of 0\farcs01 or better, and the rest of the calibrators have
position accuracies of $\sim$0\farcs1.  Multiple observations of 3C 147 and 3C
286 were used to tie the absolute fluxes to the VLA scale.

\subsection{Data Reduction} \label{data_red}

The data were processed with the Astronomical Image Processing System (AIPS; 
version 31DEC03) package.  We used R. Perley's revised coefficients for flux 
density calibration (Baars et al. 
1977) as described in the AIPS Cookbook.  Errors in the quoted flux densities 
are dominated by the uncertainty in setting the absolute flux density scale, 
with the conservative estimate being 5\%.  The data were Fourier transformed 
using AIPS task IMAGR with uniform weighting in order to maximise resolution, 
and after CLEAN deconvolution, the maps were restored.  Since we do not know 
the morphological details of the sources, we mapped the primary beam area
(5\farcm3) with 
a cell size of 0\farcs1.  The mapped region is larger than the typical size of 
the extended sources mentioned in Zakamska et~al. (2003).  The number of CLEAN 
iterations was set so that the minimum clean component reached 
0.04 mJy~beam$^{-1}$, $\sim$2 times the theoretical thermal noise limit of the 
maps.  Stronger sources (such as SDSS\,J0329$+$0052 or SDSS\,J2354$-$0056), 
whose maps were not limited by the system noise, were subjected to several 
cycles of phase-only self-calibration until the sensitivity approached thermal 
noise levels.  A model based on the CLEAN components was used to start the 
self-calibration process.  We also performed phase-only self-calibration for 
one or two iterations for the weakly detected sources, but the improvement was 
minimal.  At each round of self-calibration, the image and the visibilities 
were compared to check for the improvement in the source model.  The two IFs
were calibrated and mapped separately with the intention of concatenating
the two sets of data for each source.  The IF1 images suffered consistently 
from higher noise levels, however, and it was therefore decided to use the maps 
from IF2 (8.4601 GHz) only.  The self-calibration typically moved source 
positions by less than 0\farcs01, indicating that atmospheric phase 
irregularities had a relatively small effect on source positions.  Therefore, 
in most cases, the measured positions are accurate to 0\farcs1 or better.  The 
theoretical rms noise of our snapshot images is expected to be 
0.02 mJy~beam$^{-1}$, a level attained in almost all of the maps.

We constructed images at two different resolutions by applying appropriate
tapering functions to the visibilities.  Two sets of total intensity
(Stokes $I$) maps were made, one at full resolution (Figure~2) and
another with a Gaussian tapering function falling to 20\% of full resolution
(Figure~3), which (nearly) corresponds to a synthesized beam matching the FIRST 
survey maps.  The latter tapered maps were used to assess the possible 
presence of extended emission, to evaluate the spectral indices, and to examine 
structural counterparts between the 8.4~GHz and 1.4~GHz maps.

\section{Results}

\subsection{Maps and Source Parameters}

Our results are shown in Figure~2 for all 59 sources, including 24 
nondetections.  Sources are arranged in the order of increasing RA.  The 
restoring beam is depicted as an ellipse on the lower left-hand corner of each 
map.  The contour levels of the maps are rms $\times$ ($-$3, 3, 6, 12, 24, 48, 
...), where the rms values for the maps are given in Table~2 (Column 5).  The 
optical position of the galaxy is marked with a cross, the semi-major length of 
which corresponds to the 3 $\sigma$ uncertainty of $\sim$0\farcs3 for the SDSS 
astrometry (Pier et al. 2003).

The dynamic range (peak/noise) of the maps lies between 
$\sim 3$ and 465.  We have adopted a consistent method in extracting source 
parameters. The rms noise of each map is determined from a source-free, 
rectangular region. The galaxy is considered detected if a source with a peak 
flux density at least 5 times the rms is found within the error box of the 
optical position. For undetected sources, the upper limit is set to 5 times the
rms. Whenever possible we determined the source parameters --- integrated flux 
density and deconvolved (half maximum) sizes (major and minor axis) --- by 
fitting a two-dimensional Gaussian model using the AIPS task JMFIT.
This procedure works well for sources with relatively simple, symmetric
structure, such as most galaxy cores and pointlike features. For components
with more complex morphologies, or for weak, marginal detections, we simply
integrated the signal within interactively defined boundaries, using the AIPS
task IMEAN for rectangular boxes and TVSTAT for irregularly shaped regions. 
A source is considered resolved if its
deconvolved size is larger than one-half the beam size at full width at
half-maximum (FWHM) in at least one of the two dimensions. 

Table 2 lists the map and source parameters.  The columns are as follows:  (1) 
SDSS galaxy name, which also encodes the optical position of the source; (2) 
FWHM of the elliptical Gaussian restoring beam; (3) P.A. of the restoring beam; 
(4) flux density of the source at 8.4 GHz as measured in the default, 
high-resolution map; (5) rms noise of the map; and (6) comments on the source 
structure.  We have also gathered 1.4 GHz data from FIRST (Columns 7 and 8).  
The typical detection threshold is 
0.75 mJy, and when neither a FIRST nor an NVSS (Condon et al. 1998) detection 
is available, we use this threshold as the upper limit.  Columns (9) and (10) 
list the corresponding parameters for the tapered maps at 8.4 GHz, matched to 
the resolution of FIRST.  Finally, Column (11) gives the spectral index 
between 1.4 GHz and 8.4 GHz for the core and detected components; we have not 
assigned formal error bars to the measurements, but the uncertainty for a 
typical 1--5 mJy source is 0.04--0.07.  The low-resolution maps have restoring 
beams of size 5\farcs4$\times$5\farcs4 for sources with declination 
$\delta$ $>$30\deg\ and 6\farcs4$\times$5\farcs4 for sources with 
declination $-5$\deg\ $<$ $\delta$ $<$ 10\deg\ (an exception is
SDSS\,J0257$-$0632, which has a beam of 6\farcs8$\times$5\farcs4).

A compilation of derived radio parameters, along with optical parameters, is 
given in Table~3.   The columns are as follows: (1) galaxy name; (2) redshift, 
as determined by SDSS; (3) luminosity distance; (4) total, integrated 
monochromatic power at 1.4 GHz and (5) 8.4 GHz; and (6) the radio morphology of 
the sources shown in the maps presented.  We adopt the definitions of Ho \& 
Ulvestad (2001) and Ulvestad \& Wilson (1984) for the radio morphology classes: 
``U'' (single, unresolved), ``S'' (single, slightly resolved), ``D'' (diffuse), 
``L'' (linear structure, or multiply aligned components), and ``E'' (extended 
diffuse).  Following Ho \& Ulvestad (2001), a slightly resolved source is one 
whose deconvolved size is $\ge$ $\frac{1}{2}$ the synthesized beam width.
Radio powers in the source frame that were emitted in our observed
frequencies were computed from the flux densities using Ned Wright's
online cosmology calculator\footnote{{\tt http://www.astro.ucla.edu/$\sim$wright/CosmoCalc.html}}.  The radio powers $P^{\rm tot}_{1.4}$ and $P^{\rm tot}_{8.4}$
have been $K$-corrected to the restframe of each source, such that 
$P_{\nu} = 4 \pi D_L^2 S_\nu (1+z)^{-1-\alpha}$, with $D_L$ being
the luminosity distance, $S_\nu$ the observed flux density, and $\alpha$ the 
spectral index.  For the undetected sources, we adopt the median spectral 
index of the detected sources ($\alpha^{8.4}_{1.4} = +0.094$; see below).  The 
last two columns of the table give the luminosity of the (7) \oii\ 
$\lambda$3727 and (8) \oiii\ $\lambda$5007 lines, taken from Zakamska et al. 
(2003).  Additional notes on some selected objects are given in the 
Appendix.

\section{Discussion}

\subsection{Basic Radio Properties}

To a depth of rms $\approx$ 20 $\mu$Jy beam$^{-1}$, the detection rate of our 
8.4 GHz survey is 59\% (35/59).  Remarkably, this detection rate is essentially 
identical to that obtained from FIRST at 1.4 GHz, which observed 56 out of the 
59 sources in our sample and detected 35 (63\%), even though the sensitivity of 
FIRST is nearly an order of magnitude lower than that of our X-band survey.  
Moreover, our 8.4 GHz observations did not yield any new detections that were 
not already contained in FIRST.  The detection rate in FIRST for our subsample 
of type~2 quasars agrees well with the overall detection rate reported for the 
original SDSS sample from Zakamska et al. (2003) (46\%), as well as for the 
larger, updated sample from Reyes et al. (2008) (61\%).  Restricting the Reyes 
et al. sample to the 86 objects that meet the same selection criteria employed 
in this study ($z> 0.3$ and $L_{\rm [O~\sc III]} = 10^{41.8}-10^{42.1}$ \lum), 
the detection rate in FIRST becomes 53\%, again compatible with ours.

At our arcsecond-scale resolution, the morphology of the radio emission is 
predominantly that of a compact core, either unresolved or slightly resolved 
(75\% of the detections classified as ``U'' or ``S'').  At an average redshift
of $z=0.45$, our 0\farcs8 beam corresponds to a physical diameter of 4.6 kpc.
Higher-resolution observations may begin to resolve this emission into 
subgalactic-scale structures with physical dimensions below 1 kpc, as often 
seen in nearby Seyfert galaxies (e.g., Ho \& Ulvestad 2001).  Six sources 
($\sim$10\%) have very extended, supergalactic-scale structures (see Figure~3).
The maximum linear extents are enormous, five out of the six having diameters 
approaching 1 Mpc, comparable to the sizes of the largest radio galaxies (e.g., 
Machalski \& Jamrozy 2006).  Down to the 3 $\sigma$ contour level of the 1.4 GHz
maps, the following are the maximum linear extents: 177 kpc (SDSS~J0040$-$0040),
868 kpc (SDSS~J0257$-$0632), 709 kpc (SDSS~J0902$+$5459), 847 kpc 
(SDSS~J1008$+$4613), 797 kpc (SDSS~J1247$+$0152), and 894 kpc 
(SDSS~J2157$+$0037).  All but SDSS~J1247$+$0152 have steep spectra (average 
$\alpha^{8.4}_{1.4} = -0.87$), the latter being relatively flatter with 
$\alpha^{8.4}_{1.4} = -0.41$.

Normal, inactive $L^*$ galaxies have a characteristic radio power of $\sim 
10^{28}$ \lum\ \perhz\ at 1.4 GHz (Condon 1992; de~Vries et al. 2007).  Our 
sample of type~2 quasars is roughly 2--4.5 orders of magnitude more powerful 
(Figure 4).  Accounting for the upper limits using the Kaplan-Meier 
product-limit estimator (Feigelson \& Nelson 1985), the mean radio power is 
$7.29 \times 10^{31}$ \lum\ \perhz\ at 1.4 GHz and $1.99 \times 10^{31}$ \lum\ 
\perhz\ at 8.4 GHz; the corresponding medians are $1.69 \times 10^{30}$ and 
$3.67 \times 10^{30}$ \lum\ \perhz.  It is interesting to note that the 
distribution of radio powers at 8.4 GHz shows an apparent gap between the bulk 
of the detections and the upper limits: there appears to be an absence of 
points near $P^{\rm tot}_{8.4} \approx 10^{30}$ \lum\ \perhz.  This could 
indicate that the detected and undetected sources define two distinctly 
different populations.

By combining the 1.4 GHz FIRST measurements with our tapered, matched-resolution
8.4 GHz maps, we are able to determine spectral indices for all the detected 
components.  Because the two bands were not observed contemporaneously, we 
caution that the spectral index for any individual source might be affected by 
variability.  We do not know whether type~2 quasars vary in the radio, but if 
they are similar to type~1 quasars (Barvainis et al. 1996, 2005), they 
should.  The statistical ensemble properties of the sample, however, should
be quite robust to the effects of individual source variability.  A surprising 
aspect of our results is the preponderance of sources with rather flat to 
highly inverted spectral indices (Figure 5).  For the subset of detected 
sources, $\alpha^{8.4}_{1.4}$ has an average value of 
$-0.10$, a standard deviation of 0.65, and a median value of $+0.094$.  
Twenty-three out of the 36 (64\%) sources with spectral index measurements have 
$\alpha^{8.4}_{1.4} > -0.5$, a commonly used definition for a flat spectrum.  
Curiously, the distribution of spectral indices appears bimodal: there is a 
peak centered at $\alpha^{8.4}_{1.4} \approx -0.7$, one centered at $\sim +0.6$,
and an apparent minimum near $\alpha^{8.4}_{1.4} = 0$.  Based on a small number 
of objects with available multifrequency measurements, Zakamska et al. (2004) 
concluded that type~2 quasars tend to have relatively steep radio spectra, most 
with $\alpha$ \lax\ $-0.5$.  This result, however, is biased toward the 
minority of objects that had low-resolution literature data, most being 
powerful, steep-spectrum sources.  In a similar vein, the population of 
high-redshift ($z \approx 2$) type~2 quasars studied by Mart\'\i nez-Sansigre 
et al. (2006) is dominated by relatively powerful, steep-spectrum ($\alpha 
\approx -1$) sources. By contrast, our study, concentrating on the more typical 
members of the population, indicates the opposite --- {\it the majority of 
optically selected type~2 quasars have flat radio spectra}.  Plotting the 1.4 
GHz power as a function of spectral index (Figure~6) reveals the interesting 
trend that the spectral index becomes progressively flatter with decreasing 
radio power.  Rather than a tight correlation, the trend can best be described 
as an upper envelope: when $\alpha$ is steep, the radio power spans a very wide 
range (in this sample, up to 3 orders of magnitude), but as $\alpha$ flattens, 
the range of radio power narrows systematically, until it converges to a 
$P^{\rm tot}_{1.4} \approx 10^{30}$ \lum\ \perhz\ with a remarkably tiny spread 
($\sim$0.2 dex) when $\alpha^{8.4}_{1.4}$ approaches $+0.9$.  If we focus 
on the population with $P^{\rm tot}_{1.4}$ \lax\ $10^{31}$ \lum\ \perhz, 
close to the traditional luminosity-based criterion for radio-quiet AGNs
(e.g., Miller et al. 1990; Hooper et al. 1995), essentially {\it all sources 
are flat-spectrum.}

Most of the sources with $\alpha^{8.4}_{1.4} \ge 0$ have a compact radio 
structure, but, interestingly, not all compact sources have a flat or inverted 
spectrum.  SDSS~J0316$-$0059, J0903$+$0211, 
J0908$+$4347, J1539$+$5142, and J2354$-$0056 all have compact structures but 
are steep-spectrum; they are also moderately radio-loud (see Section 4.2). 
These may be analogs of compact steep-spectrum sources (Fanti et al. 1990), 
although classical compact steep-spectrum sources are 2--3 orders of magnitude 
more luminous than our objects (O'Dea 1998).  Conversely, while extended, 
linear features are typically steep-spectrum (e.g., SDSS~J0257$-$0632, 
J0902$+$5459, J1008$+$4613), there are some examples of double-lobed structures 
that have clearly flat or inverted spectra (e.g., SDSS~J0741$+$3020, 
J1014$+$0244).  This is unusual, but not without precedence (e.g., Lal \& Rao 
2007).

Comparable statistics for the spectral indices of radio-quiet type~1 quasars 
in the GHz range are fragmentary and highly incomplete. Nevertheless, to get 
a rough comparison, we compiled available measurements for low-redshift 
radio-quiet quasars, drawn from the studies of Barvainis et al. (1996, 2005), 
Kukula et al. (1998), and Ulvestad et al. (2005).  Similar to our observations,
all of these measurements were made with the VLA, and the frequency coverage
was from 1.4 GHz to either 5 or 8.4 GHz.  For a heterogeneous collection of 
73 objects, we find a mean spectral index of $-0.46$ and a median value of 
$-0.59$; 51\% qualify as flat-spectrum ($\alpha > -0.5$).  Falcke et al. (1996)
conducted a quasi-simultaneous single-dish survey at 2.7 and 10 GHz of all
Palomar-Green (Schmidt \& Green 1983) quasars with 4.9 GHz flux densities 
greater than 4 mJy.  For the 25 radio-quiet sources (those with $R < 10$ 
as defined by Kellermann et al. 1989) with tabulated spectral indices, we 
find a mean value of $-0.66$ and a median value of $-0.60$; 28\% of these 
qualify as flat-spectrum.  Although a more rigorous control sample of type~1 
quasars would clearly be desirable, the above comparison suggests that 
flat-spectrum sources are much more prevalent in type~2 quasars than in type~1 
quasars.  Taken at face value, this implies that the two classes are {\it not}\
intrinsically the same, in apparent violation of the simplest notions of the 
orientation-based unified model.

\subsection{The Nature of the Radio-Intermediate, Flat-Spectrum Sources}

The radio output of type~1 quasars (e.g., Kellermann et al. 1989) and lower 
luminosity Seyfert 1 nuclei (e.g., Ho \& Peng 2001) is conventionally specified 
by referencing the radio band with respect to the broadband optical continuum.
This definition of ``radio-loudness'' cannot be used for obscured, type~2 
sources, but the luminosity of the largely isotropic narrow emission lines can 
be used instead to substitute for the optical continuum.  Figure~7 illustrates 
the distribution of integrated radio power (at 5 GHz) versus \oiii\ luminosity 
for a large, heterogeneous sample of AGNs compiled by Xu et al. (1999).  The 
points clearly delineate two nearly parallel sequences, albeit each with 
significant scatter, separated by a pronounced gap.  Roughly $15\%\pm5\%$ of
our sample lie on the radio-loud branch.  Within the somewhat nebulous
definition of this nomenclature and the small statistics of our sample, this is
consistent with the radio-loud fraction ($9\%\pm2\%$) quoted by Zakamska et al.
(2004) for the parent SDSS sample, which in turn is consistent with the nominal
radio-loud fraction of type~1 quasars ($\sim 10\%-20\%$; Kellermann et al.
1989; Hooper et al. 1995; Ivezi\'c et al. 2002).  Not surprisingly, all the
nondetections are in the radio-quiet branch.  Figure~8 gives a magnified view
of our sample alone.   As already noted above, all but one of the 8.4 GHz upper
limits seem to define a population separate from the detected objects; there
appears to be a genuine dearth of points near $P^{\rm tot}_{8.4} \approx
10^{30}$ \lum\ \perhz.

Within the Xu et al. (1999) sample, a small number of objects (3\%; marked as
green diamonds), sometimes called ``radio-intermediates,'' straddle the 
radio-loud/radio-quiet divide.  Among type~1 quasars, Miller et al. (1993) and 
Falcke et al. (1996) argue that relativistic boosting of the radio-quiet 
population, with modest Lorentz factors of 2--4, can account for the 
radio-intermediate objects.  An important implication of this interpretation 
is that radio-quiet 
quasars, as in radio-loud quasars and radio galaxies, also possess relativistic 
jets.  This hypothesis offers a natural explanation for the tendency of these 
sources to be flat-spectrum, compact, and variable (Miller et al. 1993; Falcke 
et al. 1996; Kukula et al. 1998; Barvainis et al. 2005; Wang et al. 2006).  
When available, VLBI-scale imaging confirms the high brightness temperatures 
expected from partially opaque synchrotron cores (Falcke et al. 1997; Blundell 
\& Beasley 1998; Ulvestad et al. 2005), as seen in classical radio-loud sources.

Intriguingly, {\it the vast majority of the detected sources in our sample fall 
in the gap occupied by radio-intermediate sources.}\footnote{Glikman et al. 
(2007) note a similar trend for red quasars.}  As in radio-intermediate 
type~1 quasars, the type~2 counterparts are largely compact sources, and they, 
too, have flat or inverted spectra.  Despite these superficial similarities, 
however, the large number of radio-intermediate sources relative to the 
radio-quiet ones makes it highly improbable that all of the former are 
Doppler-boosted versions of the latter.  It would appear that at least another 
mechanism is needed to explain the flatness of the radio spectra.  

As in type~2 quasars, lower luminosity Seyfert galaxies (both type~1 and 2) 
also frequently possess flat-spectrum cores (Ho \& Ulvestad 2001), whose 
physical origin is not well understood.  Apart from self-absorbed synchrotron 
emission from a compact core\footnote{Laor \& Behar (2008) propose the novel 
idea that compact, flat-spectrum cores in radio-quiet (and presumably also 
radio-intermediate) AGNs may arise from a magnetically heated corona associated
with the accretion disk.}, Ulvestad \& Ho (2001) suggest that free-free 
absorption by thermal, ionized gas in the vicinity of the nucleus might be 
sufficient to flatten intrinsically steeper synchrotron spectra.  The same 
argument can be plausibly extended to the case of the type~2 quasars under 
consideration.  Following the exercise outlined in Section 4.4.3 
of Ulvestad \& Ho (2001), the emission measure of a $10^4$ K gas needed to 
achieve a free-free optical depth of unity at a turnover frequency of, say, 5 
GHz, is $E \approx  9 \times 10^7\,{\rm cm}^{-6}\,{\rm pc}$.  If the gas has an 
electron density of $10^3$ \cc, a value typical of the narrow-line region, the 
required path length is $\sim 90$ pc.  Under case B recombination and a 
filling factor of unity, a spherical volume of this radius would produce 
an H\al\ luminosity of $8\times10^{43}$ \lum.  This compares favorably with 
the strengths of the narrow emission lines observed in our sample.  Recall 
that our sources were picked to have $L_{\rm [O~{\sc III}]} \approx 10^{42}$ 
\lum.  From the composite spectrum tabulated in Zakamska et al. (2003), we find 
that \oiii/H$\beta$ = 5.5 and H\al/H$\beta$ = 4.  Assuming an intrinsic 
Balmer decrement of H\al/H$\beta$ = 3.1 and the Galactic extinction law of 
Cardelli et al. (1989), the 
inferred H\al\ luminosity is $\sim 1.3\times10^{42}$ \lum.  The observed 
and predicted values of the H\al\ luminosity can be easily reconciled by 
invoking a modest filling factor of $\sim 10^{-2}$.  This back-of-the-envelope 
calculation is only meant to be suggestive, but it does serve to illustrate
that the typical properties of the narrow-line region in principle can
impart sufficient free-free absorption to flatten an intrinsically steep 
radio spectrum.  In this picture, the preponderance of flat-spectrum sources 
in type~2 quasars relative to type~1 quasars can then be attributed to 
differences in the detailed structure or nebular properties of their 
narrow-line regions.  This may not entirely unexpected, in light of the 
evidence already emerging that the host galaxies of type~2 quasars may
have experienced recent mergers or tidal interactions (Greene et al. 2009; 
Liu et al. 2009) and that their narrow-line regions are dynamically unrelaxed 
(Greene et al. 2009).  Precisely how these conditions result in enhanced 
free-free absorption toward the nucleus, however, is unknown.

\subsection{Contribution from Star Formation}

As noted by Kim et al. (2006), the host galaxies of type~2 quasars, unlike 
those of type~1 objects (Ho 2005), do show significant levels of ongoing star 
formation activity, as evidenced by their elevated \oii\ \lamb3727 emission.  
Comparing the observed emission-line ratios with photoionization models, 
Villar-Mart\'\i n et al.  (2008) reached a similar conclusion, finding that 
young stars contribute partially to the excitation of the SDSS sources, 
especially ones with lower \oiii\ luminosities, such as those studied here.  
However, as we now demonstrate, the expected level of star formation falls far 
short of that needed to account for the radio emission observed.  From the 
optical line luminosities listed in Table~3, the \oii\ \lamb3727/\oiii\ 
\lamb5007 ratio has a median value of 0.45.  This is a factor of 2--3 higher 
than expected from the narrow-line regions of high-ionization AGNs such as 
Seyfert galaxies and quasars (Ho 2005), strongly suggesting that \hii\ regions 
do indeed contribute to the integrated optical line emission in these sources.
But the inferred star formation rates (SFRs) are modest.  Following Ho (2005), 
we estimate that the sources in our sample have \oii-derived SFRs of only 
2--15 $M_{\sun}\, {\rm yr^{-1}}$, with a median value of $\sim 5$ $M_{\sun}\, 
{\rm yr^{-1}}$.  We can estimate the corresponding level of radio emission 
from empirical SFR estimators.  The prescription of Bell (2003; Equation 6), 
SFR$(M_{\sun}\, {\rm yr^{-1}}$)= $5.52 \times 10^{-29} P_{\rm 1.4}$, predicts a 
radio power of merely $9\times 10^{28}$ \lum\ \perhz\ at 1.4 GHz, or, for 
$\alpha = -0.7$, $\sim 3\times 10^{28}$ \lum\ \perhz\ at 8.4 GHz, which is 
about an order of magnitude below the upper limits of our present observations 
(Figure 8).  The contribution of star formation and host galaxy emission to the 
current radio measurements is therefore completely negligible.  

The radio-intermediate sources in our sample, especially, cannot be explained 
as radio-quiet objects whose host galaxy radio emission has been enhanced as 
a consequence of vigorous star formation.  The flat or inverted spectra of these
sources do not permit a strong contribution from optically thin synchrotron 
emission, as would be the case if a dominant starburst component were present.  
More seriously, the typical radio power of $P^{\rm tot}_{1.4} \approx 10^{31}$ 
\lum\ \perhz\ translates into a SFR of $\sim 550$ $M_{\sun}\, {\rm yr^{-1}}$.  
While such extreme SFRs are not unheard of in high-redshift obscured quasars 
(e.g., Mainieri et al. 2005; Vignali et al. 2009), they are much higher than 
deduced even for the most luminous members of the optically selected SDSS 
sample (Zakamska et al.  2008; Liu et al. 2009). And they would be in flagrant 
violation of the observed \oii\ line strength, even after allowing for the 
vagaries of extinction correction.

\section{Summary and Future Directions}

This paper presents deep, high-resolution 8.4 GHz VLA images of a well-defined
sample of 59 optically selected type~2 quasars drawn from the SDSS sample of 
Zakamska et al. (2003).  The sample focuses on a narrow range in \oiii\ \lamb 
5007 luminosity, at $L_{\rm [O~\sc III]} \approx 10^{42}$ \lum.  Our main 
results can be summarized as follows: 

\begin{enumerate}

\item{Our detection rate at 8.4 GHz, 59\%, is essentially identical to that 
of the FIRST survey at 1.4 GHz, even though our sensitivity is nearly an order 
of magnitude deeper.}

\item{We detect a high fraction (75\%) of compact cores, which confine the 
radio emission to typical physical diameters of 5 kpc or less.  About 10\% 
of the objects contain very extended, steep-spectrum structures, most having 
physical extents approaching 1 Mpc.}

\item{Roughly 15\%$\pm$5\% of the sample have radio-to-\oiii\ luminosity ratios 
that qualify them as radio-loud sources.}

\item{Most of the detected sources have flat or inverted spectra 
between 1.4 and 8.4 GHz and radio-loudness parameters similar to 
so-called radio-intermediate quasars.  They are too numerous relative to 
the radio-quiet population to be all Doppler-boosted members of the latter.  
The incidence of flat-spectrum sources, which might arise from free-free 
absorption in the narrow-line region, is higher in type~2 quasars than in 
type~1 quasars, in apparent conflict with the orientation-based unified model.}

\item{The ongoing star formation rates of this sample are relatively 
modest ($\sim 5$ \solmass\ \peryr) and contribute negligibly to the detected
radio continuum emission.}

\end{enumerate}

Future radio observations can extend the present study in several directions.

\begin{itemize}
\item{The present sample, confined to $L_{\rm [O~\sc III]} \approx 10^{42}$ 
\lum, barely scratches the surface of the type~2 quasar population.  With the 
availability of the new SDSS sample from Reyes et al. (2008), it would be 
interesting to probe a much wider luminosity range, especially toward higher 
luminosities where follow-up observations at other wavelengths are already 
under way (e.g., Zakamska et al. 2008; Greene et al. 2009; Liu et al. 2009).}

\item{What is the true nature of the flat-spectrum sources?  We have offered 
some simple guesses, but none can be definitively tested yet.  The spectral 
measurements can be significantly improved by acquiring quasi-simultaneous 
observations to mitigate the effects of variability, and preferably at more 
than two frequencies to better define the spectral shape.  As demonstrated 
already for type~1 quasars, the compact jet hypothesis can be tested through 
variability measurements and VLBI/VLBA observations to constrain brightness 
temperatures.}

\item{Deeper observations of the current sample are needed to understand the 
nature of the nondetections.  Do these objects have AGN-like radio properties 
(e.g., flat-spectrum cores), or do they instead actually have steep-spectrum, 
diffuse emission, which might be the case if they were a separate population 
dominated by star formation?  To reach the necessary sensitivity for a 
statistically meaningful sample will likely require the upcoming EVLA.}
\end{itemize}

\acknowledgments

We thank the anonymous referee for his/her comments that improved this paper.
The VLA is operated by the US National Radio Astronomy Observatory which is
operated by Associated Universities, Inc., under cooperative agreement with
the National Science Foundation.  The National Radio Astronomy Observatory is 
a facility of the National Science Foundation operated under cooperative 
agreement by Associated Universities, Inc.  This research has made use of the 
NASA/IPAC Extragalactic Database (NED) which is operated by the Jet Propulsion 
Laboratory, California Institute of Technology, under contract with NASA.  We 
thank the staff of the VLA for carrying out these observations in their usual 
efficient manner.  D.~V.~L.  acknowledges the assistance from J. Lim and 
P.~T.~P. Ho on several occasions.  L.~C.~H. thanks Jim Ulvestad for advice 
and Nadia Zakamska for clarification on the \oiii\ luminosity conversion of 
her sample.  The work of L.~C.~H. is supported by the Carnegie Institution for
Science.

%----------------------------------------------------------------------

%APPENDIX
\clearpage
\appendix

\section{Notes on Selected Sources}

This section provides some additional information on the sources with 
extended structure.

{\sl SDSS\,J0040$-$0040} ---
This source appears extended in the FIRST image, as first noticed by Zakamska
et al. (2004), whereas in our 8.4 GHz, high-resolution and tapered maps the
source is classified as diffuse with faint extended emission.  The low-surface
brightness feature toward the north-west seen at 1.4 GHz is not detected at
8.4 GHz, possibly due to the absence of short ($u,v$) spacings.  The integrated
spectral index is $\alpha_{1.4}^{8.4}$ = $-$0.64.

{\sl SDSS\,J0257$-$0632} ---
This source has a double-lobed extended morphology at both 1.4 and 8.4 GHz.
The extent of the northern lobe seen at 1.4 GHz is larger than that in the
tapered 8.4 GHz map.  The integrated spectral index of $\alpha_{1.4}^{8.4}$ =
$-$0.71 is typical for extended radio sources.

{\sl SDSS\,J0741$+$3020} ---
The source is slightly resolved at both frequencies and has a highly inverted
integrated spectrum ($\alpha_{1.4}^{8.4}$ = $+$0.56).  Based on the brightest
contour, the source possibly shows a double structure.

{\sl SDSS\,J0818$+$3958} ---
The radio core is slightly resolved, possibly a core-jet morphology, and
has a moderately flat spectrum ($\alpha_{1.4}^{8.4}$ = $-$0.25).

{\sl SDSS\,J0902$+$5459} ---
This source appears extended in FIRST (Zakamska et al. 2004).  At 8.4 GHz, the
source shows a double-lobed structure, which corresponds to the extended
diffuse feature located $\sim$5\asec\ away from the phase-center toward the
northwest.  The lobe to the southeast is not detected at 8.4 GHz, possibly
because it has a very steep spectrum.  The integrated spectral index of
$\alpha_{1.4}^{8.4}$ = $-$0.89 is typical for extended radio sources.

{\sl SDSS\,J0945$+$5708} ---
The source is slightly resolved, with a possible core-jet structure, and has a
moderately flat spectrum ($\alpha_{1.4}^{8.4}$ = $-$0.39).

{\sl SDSS\,J0956$+$5735} ---
The source is slightly resolved, possibly with a core-jet morphology, but it
has a highly inverted spectrum ($\alpha_{1.4}^{8.4}$ = $+$0.72).

{\sl SDSS\,J1008$+$4613} ---
The source has a symmetric double-lobed radio counterpart, similar to
``classic'' radio galaxies, with the detected core centered on the optical
position.  This is the third source in our sample that appears clearly
extended at both 1.4 and 8.4 GHz.  The integrated spectral index is very
steep ($\alpha_{1.4}^{8.4}$ = $-$0.99).  This is the only source in our sample
detected in the X-rays (Vignali et~al. 2004).

{\sl SDSS\,J1014$+$0244} ---
The source shows double-source morphology on arcsec-scale
resolution at 8.4~GHz, whereas it is unresolved in FIRST.
The integrated spectral index is inverted, with $\alpha_{1.4}^{8.4}$ = $+$0.45.

{\sl SDSS\,J1026$-$0042} ---
The source is slightly resolved, possibly has a double-source morphology,
and has an inverted spectrum ($\alpha_{1.4}^{8.4}$ = $+$0.38).

{\sl SDSS\,J1247$+$0152} ---
This is another source in our sample that is extended in FIRST.  At 8.4 GHz,
the source appears to have a slightly resolved core with diffuse emission
around it.  We only detect the emission associated with the core component.
The integrated spectral index is mildly flat ($\alpha_{1.4}^{8.4}$ = $-$0.41).
The low-surface brightness features toward the north-west and south-east
seen in 1.4 GHz FIRST image are not detected in the low-resolution 8.4 GHz
map, possibly due to the absence of short ($u,v$) spacings.

{\sl SDSS\,J1447$+$0211} ---
The source is slightly resolved, possibly with a core-jet structure, and has
a highly inverted spectrum ($\alpha_{1.4}^{8.4}$ = $+$0.85).

{\sl SDSS\,J1548$+$0046} ---
The source is slightly resolved in FIRST, whereas it shows a double-source
morphology in our 8.4 GHz map.  It has a steep integrated spectrum of
$\alpha_{1.4}^{8.4}$ = $-$0.95.  We did not fit a double-Guassian model to
the source because it is insufficiently resolved in the FIRST map.

{\sl SDSS\,J2157$+$0037} ---
This is the last source in our sample that appears extended in FIRST images
(Zakamska et~al. 2004).  Our 8.4 GHz map, however, only detects the compact
emission.  The integrated spectral index is very steep, with
$\alpha_{1.4}^{8.4}$ = $-$1.10.  None of the low-surface brightness features
detected at 1.4 GHz is detected in the tapered 8.4 GHz map, possibly due to
the absence of short ($u,v$) spacings.

\clearpage
\figurenum{1}
\begin{figure}
\epsscale{1.00}
\plotone{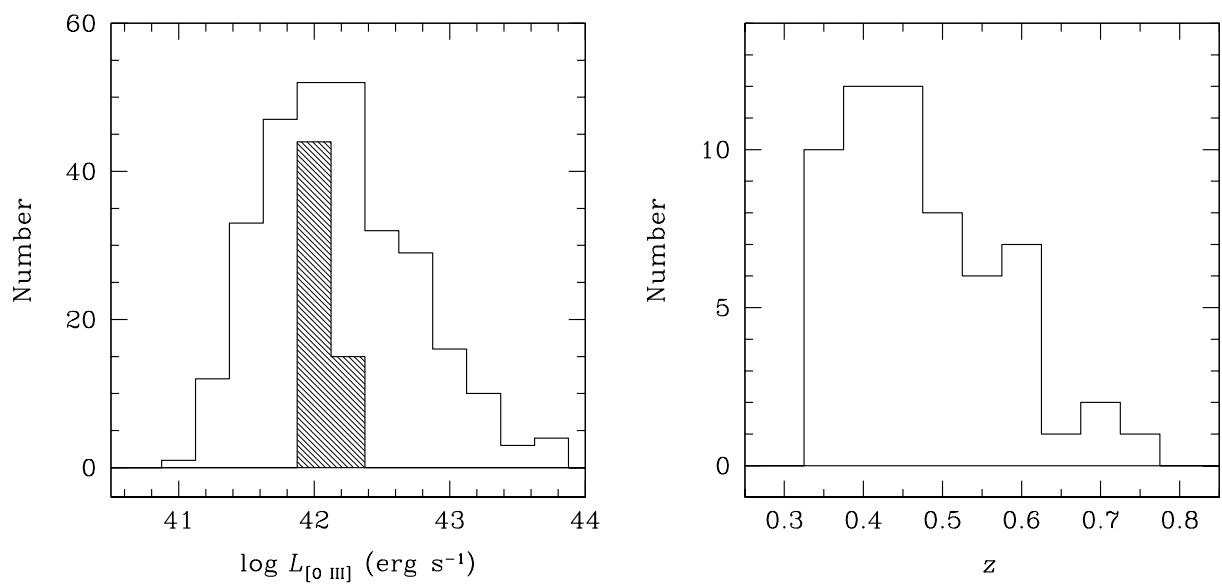}
\caption{{\it Left}: Distribution of \oiii\ $\lambda$5007 luminosity for the
parent sample of SDSS type~2 quasars from Zakamska et al. (2003; {\it solid
line}) and subsample observed with the VLA at 8.4 GHz ({\it hashed}).
{\it Right}: Distribution of redshift for the subset of SDSS type~2 quasars
observed by us.}
\label{our_samp}
\end{figure}

\clearpage
\figurenum{2}
\begin{figure}
\epsscale{0.90}
\plotone{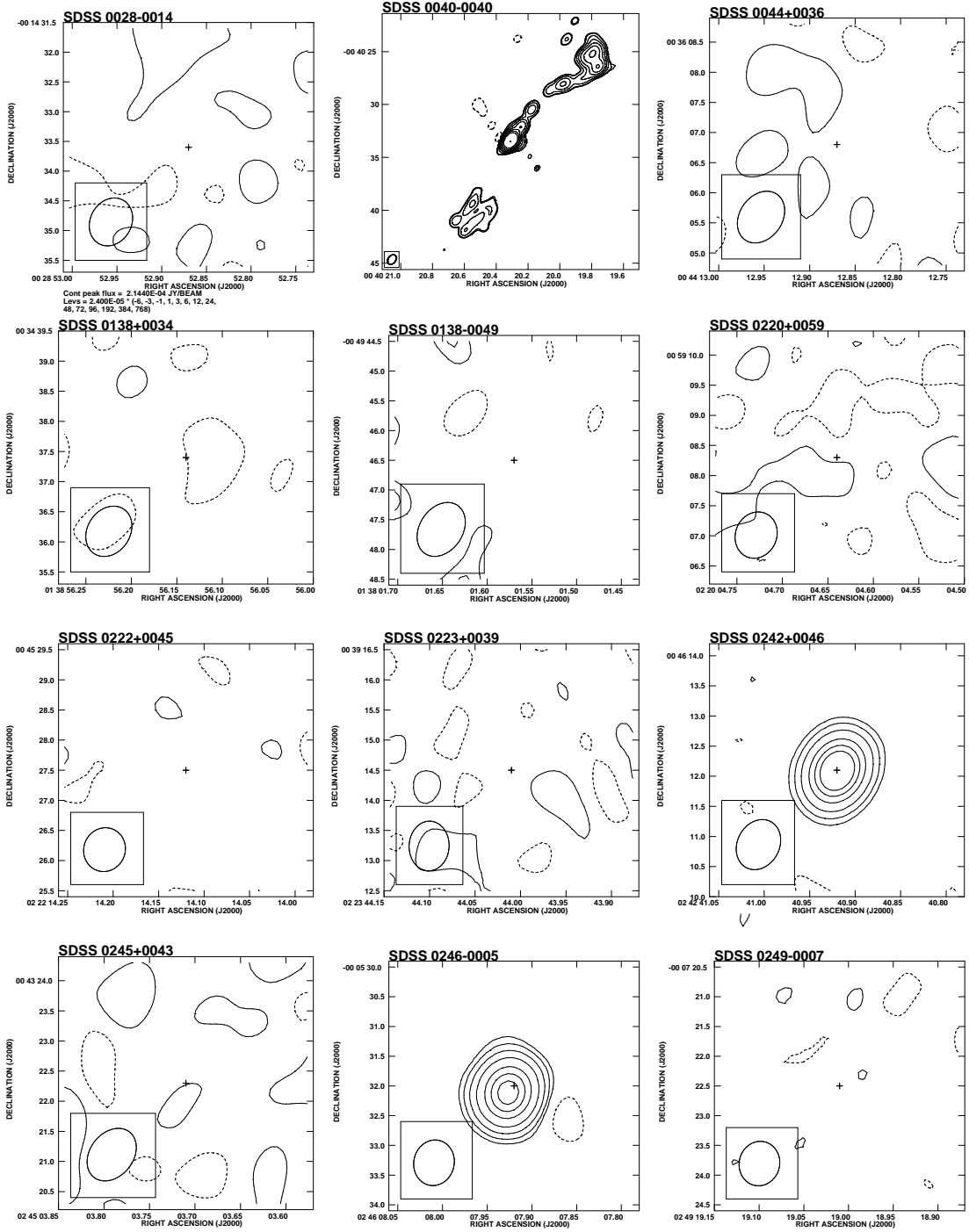}
\caption{VLA B configuration 8.4~GHz images at 0\farcs8 resolution of 
type~2 quasars. The fields are centered on the optical positions
given by Zakamska et~al. (2004), which are marked with a cross, whose
semi-major length corresponds to the 1 $\sigma$ uncertainty of 0\farcs1. Image
fields are 4\asec $\times$4\asec, except for the large angular sized source
SDSS~J0040$-$0040, for which we show 8\asec$\times$8\asec.  The restoring beam
is depicted as an ellipse on the lower-left corner of each map.  The maps
display contour levels of rms $\times$ ($-3$, 3, 6, 12, 24, 48, ...); the rms
values for the maps are listed in Table~2.}
\end{figure}

\clearpage
\figurenum{2}
\begin{figure}
\epsscale{1.00}
\plotone{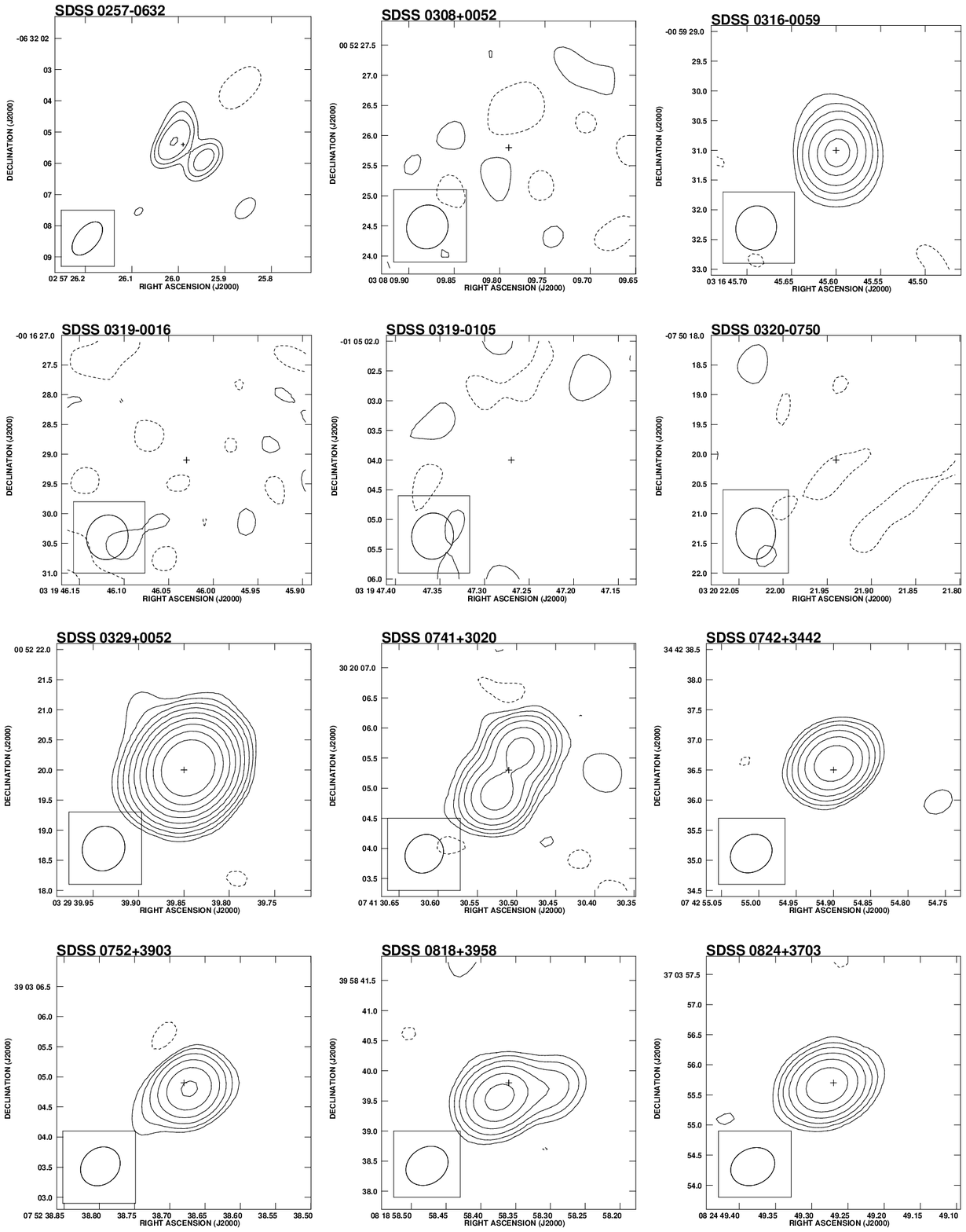}
\caption{{\it continued.}
The field for SDSS~J0257$-$0632 is 8\asec$\times$8\asec.}
\end{figure}

\clearpage
\figurenum{2}
\begin{figure}
\epsscale{1.00}
\plotone{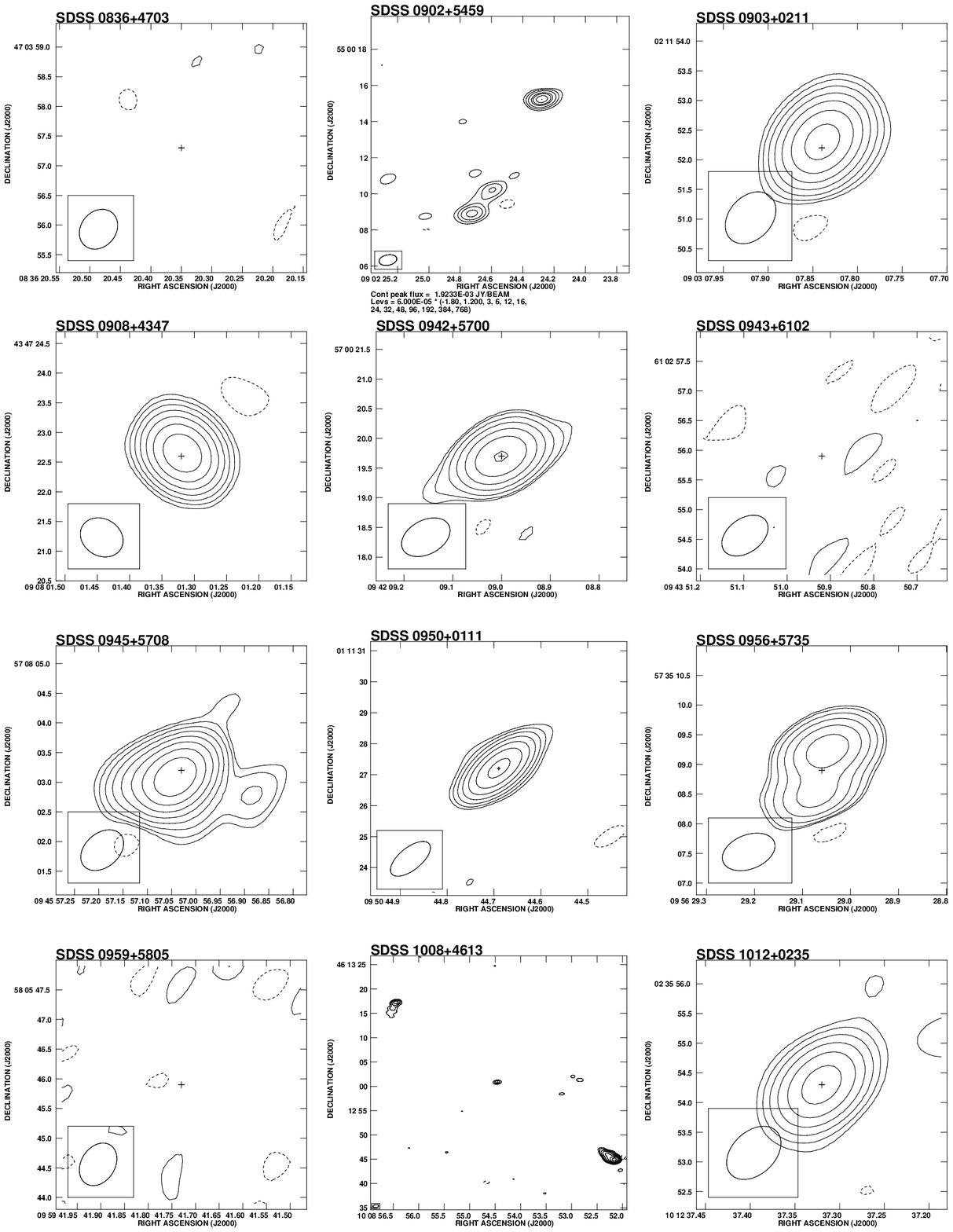}
\caption{{\it continued.}
The fields for SDSS~J0902$+$5459, J0950$+$0111, and J1008$+$4613
are 14\asec$\times$14\asec, 8\asec$\times$8\asec, and 52\asec$\times$52\asec, 
respectively.}
\end{figure}

\clearpage
\figurenum{2}
\begin{figure}
\epsscale{1.00}
\plotone{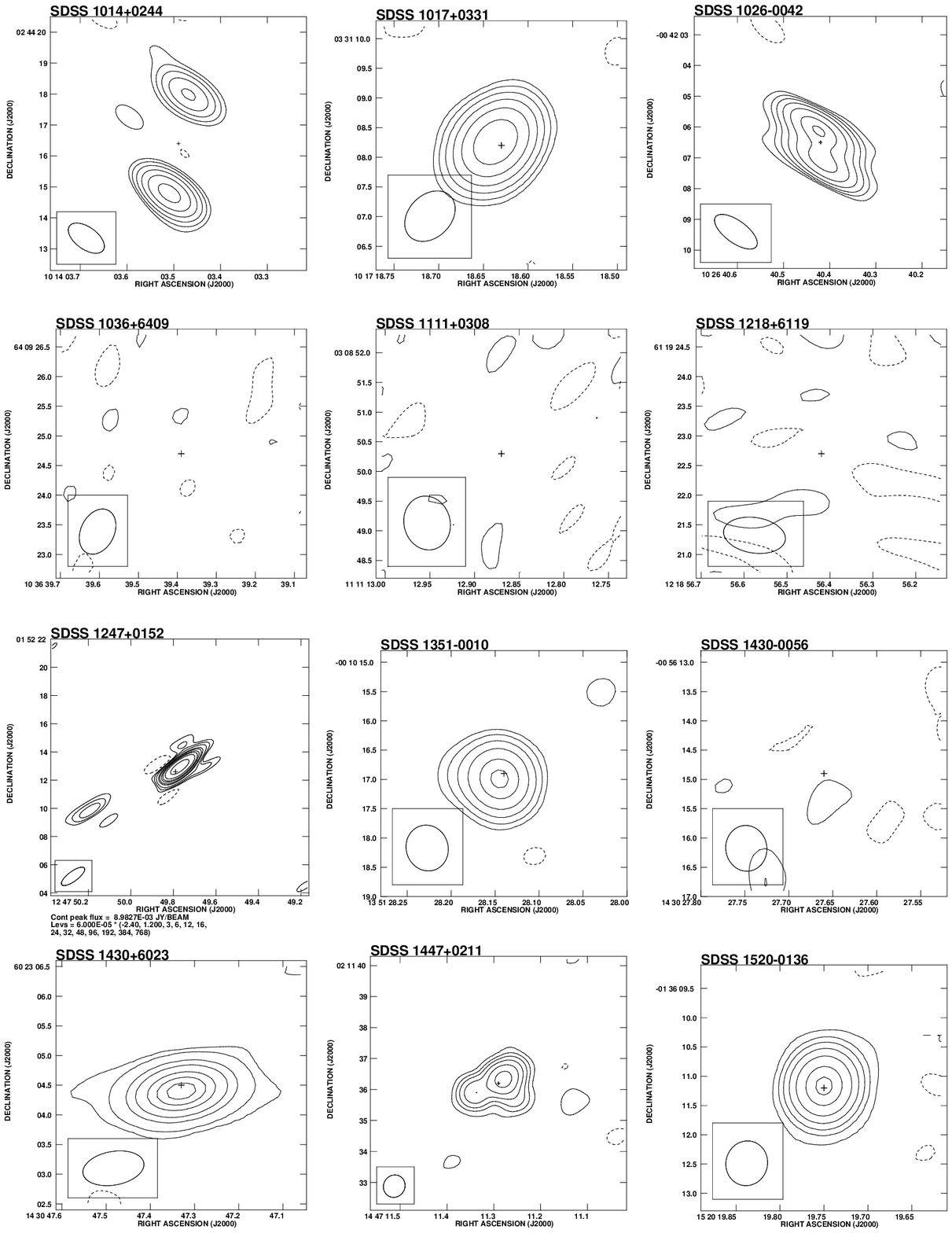}
\caption{{\it continued.}
The fields for SDSS~J1014$+$0244, J1026$-$0042, J1247$+$0152, and J1447$+$0211
are 8\asec$\times$8\asec, 8\asec$\times$8\asec, 34\asec$\times$34\asec, and 
8\asec$\times$8\asec, respectively.}
\end{figure}

\clearpage
\figurenum{2}
\begin{figure}
\epsscale{1.00}
\plotone{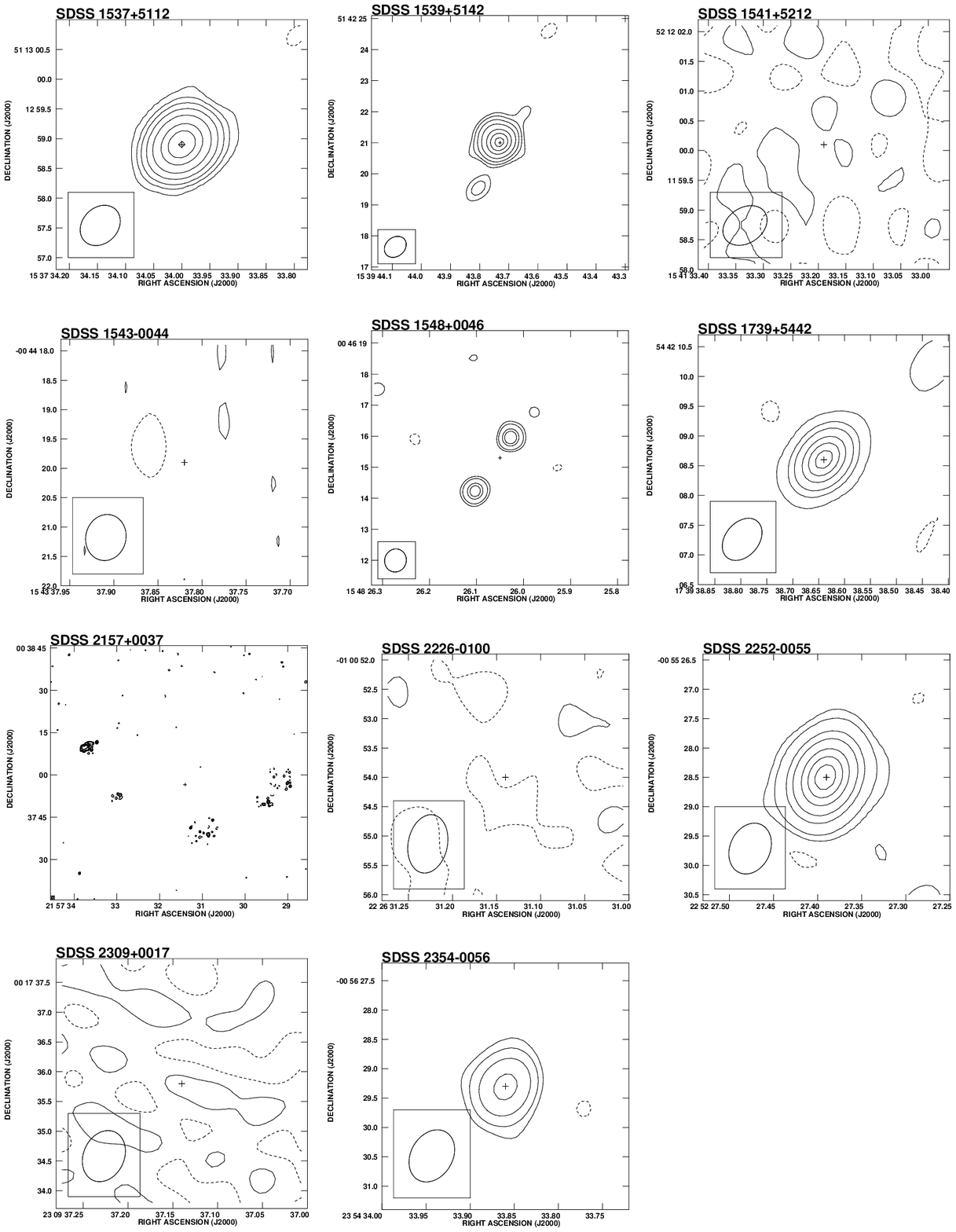}
\caption{{\it continued.}
The fields for SDSS~J1539$+$5142 and J1548$+$0046 are
8\asec$\times$8\asec, and that of J2157$+$0037 is 30\asec$\times$30\asec.}
\end{figure}

\clearpage
\figurenum{3}
\begin{figure}
\epsscale{0.93}
\plotone{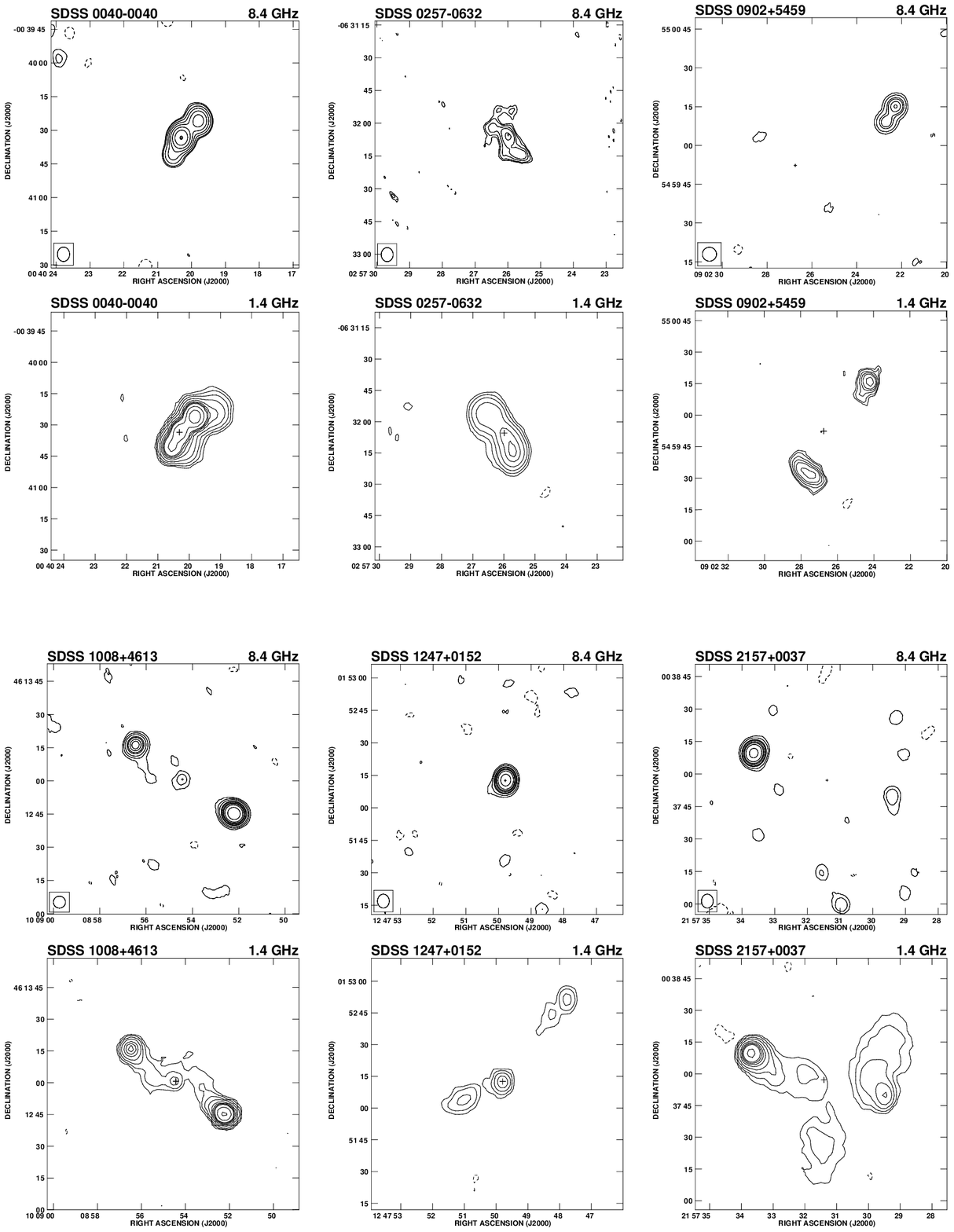}
\caption{Rows 1 and 3: Low-resolution, tapered maps of the extended
sources at 8.4 GHz matched to the resolution of the FIRST 1.4~GHz images.
Rows 2 and 4: Corresponding FIRST 1.4~GHz images at $\sim$5\asec\ resolution.
The fields are centered on the optical positions given by Zakamska et~al. 
(2004). Image fields are 2\amin$\times$2\amin. Four objects --- 
SDSSS~J0040$-$0040, J0902$+$5459, J1247$+$0152, and J2157$+$0037 --- match 
extended sources in FIRST.}
\end{figure}

\clearpage
\figurenum{4}
\begin{figure}
\epsscale{1.00}
\plotone{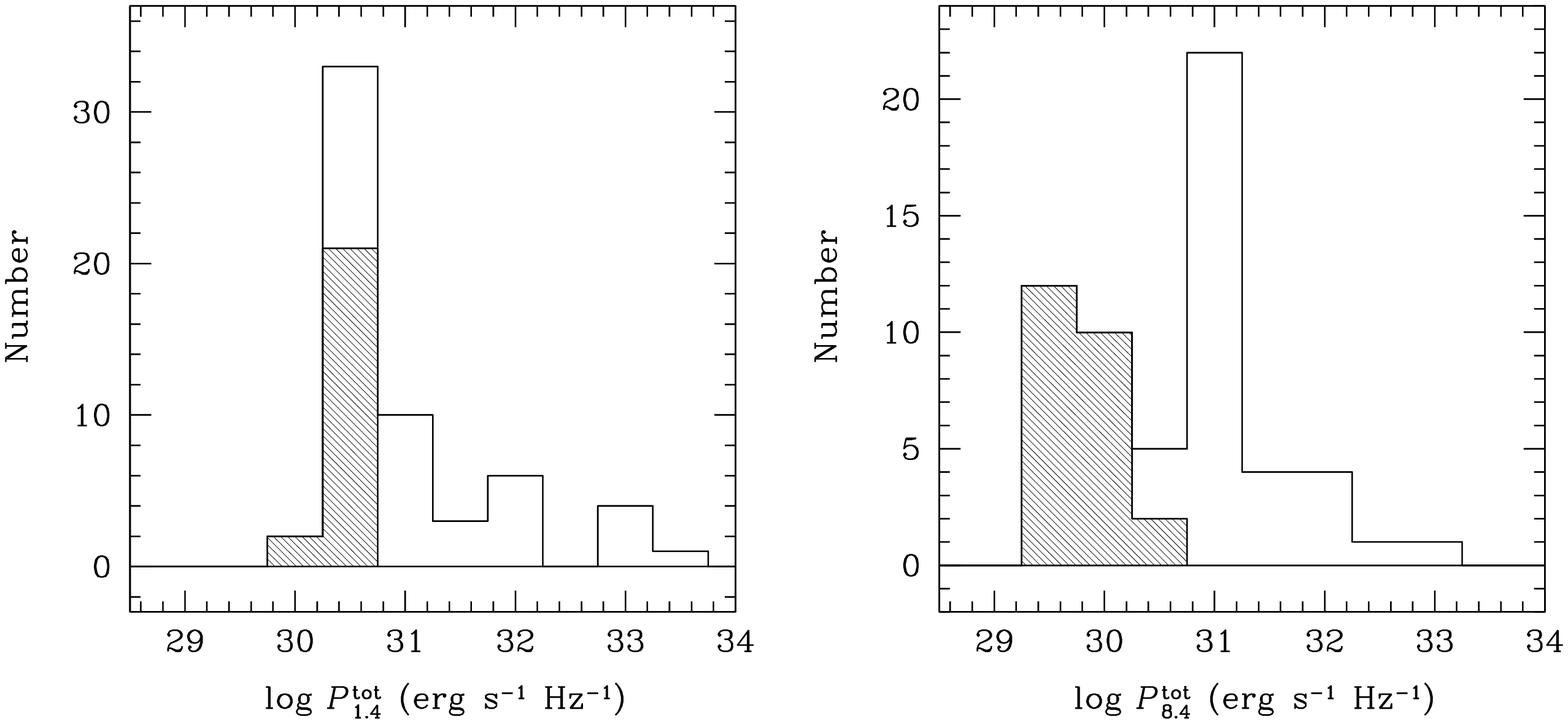}
\caption{Distribution of total integrated radio powers at 1.4 GHz ({\it left})
and 8.4 GHz ({\it right}). The shaded region of the histogram denotes upper
limits.}
\end{figure}

\clearpage
\figurenum{5}
\begin{figure}
\epsscale{1.00}
\plotone{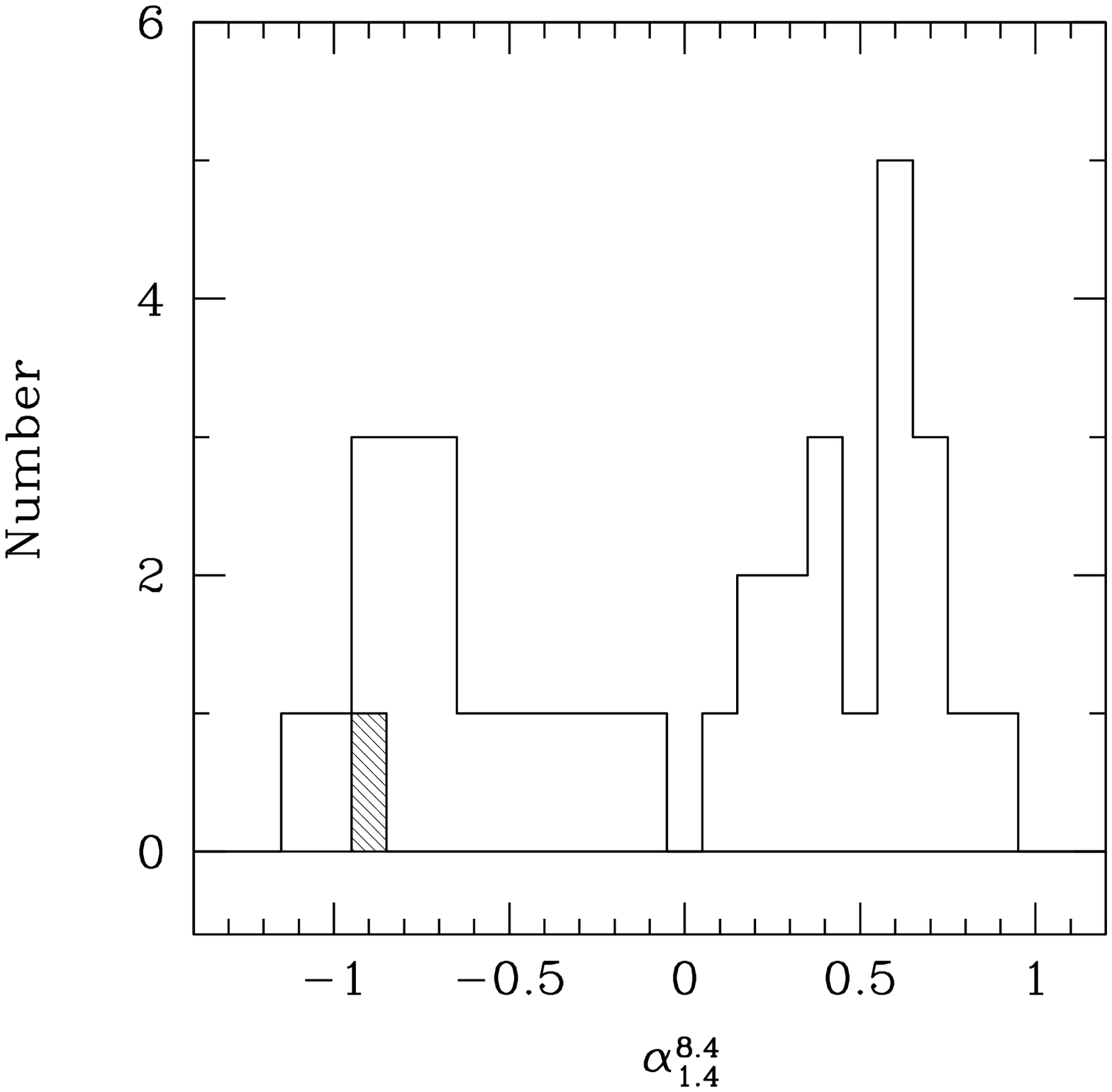}
\caption{Distribution of integrated spectral indices between 1.4 and 8.4 GHz.
The hatched histogram marks an upper limit.}
\end{figure}

\clearpage
\figurenum{6}
\begin{figure}
\epsscale{1.00}
\plotone{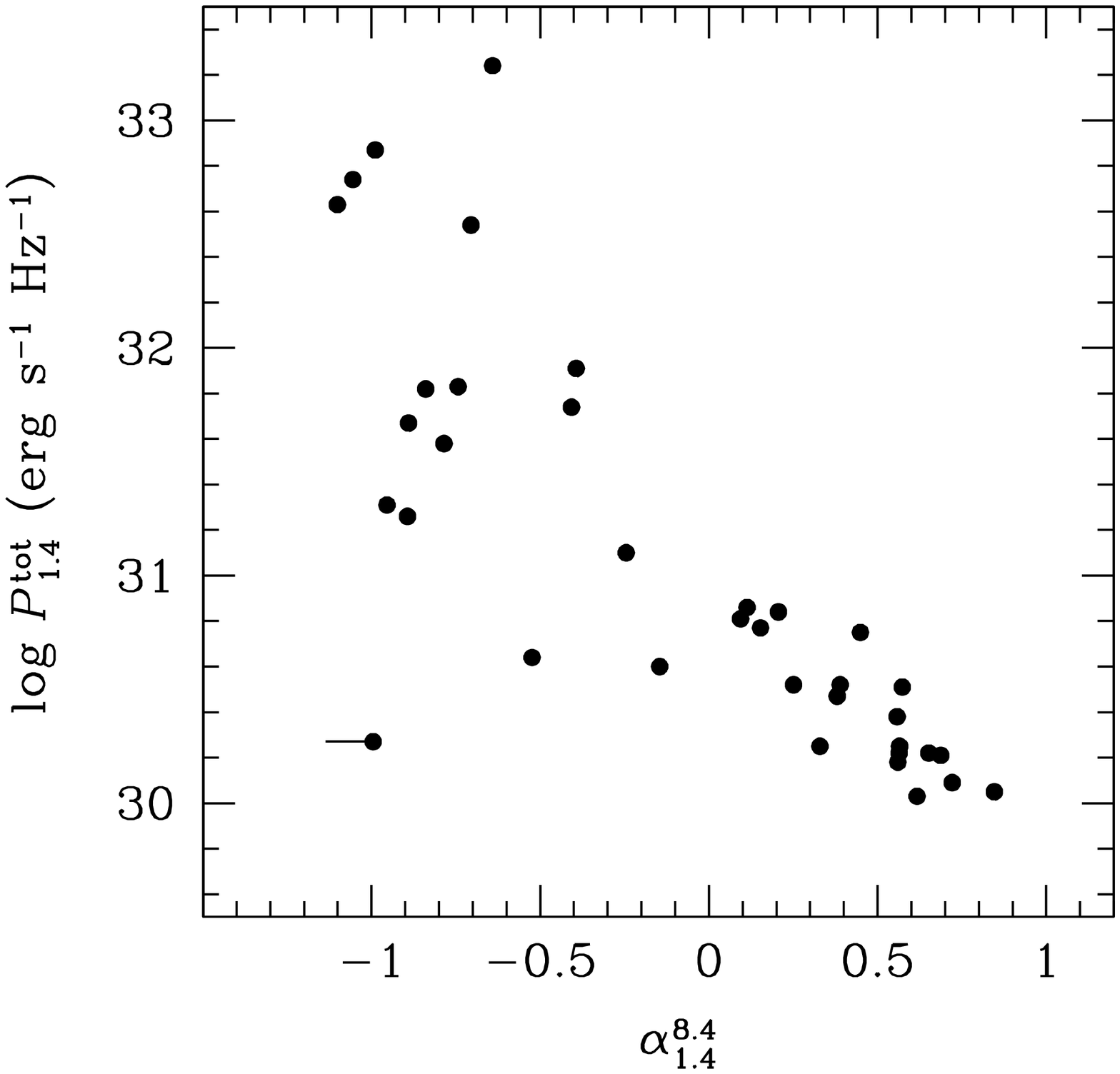}
\caption{Distribution of radio power at 1.4 GHz versus integrated spectral 
indices between 1.4 and 8.4 GHz.}
\end{figure}

\clearpage
\figurenum{7}
\begin{figure}
\epsscale{1.00}
\plotone{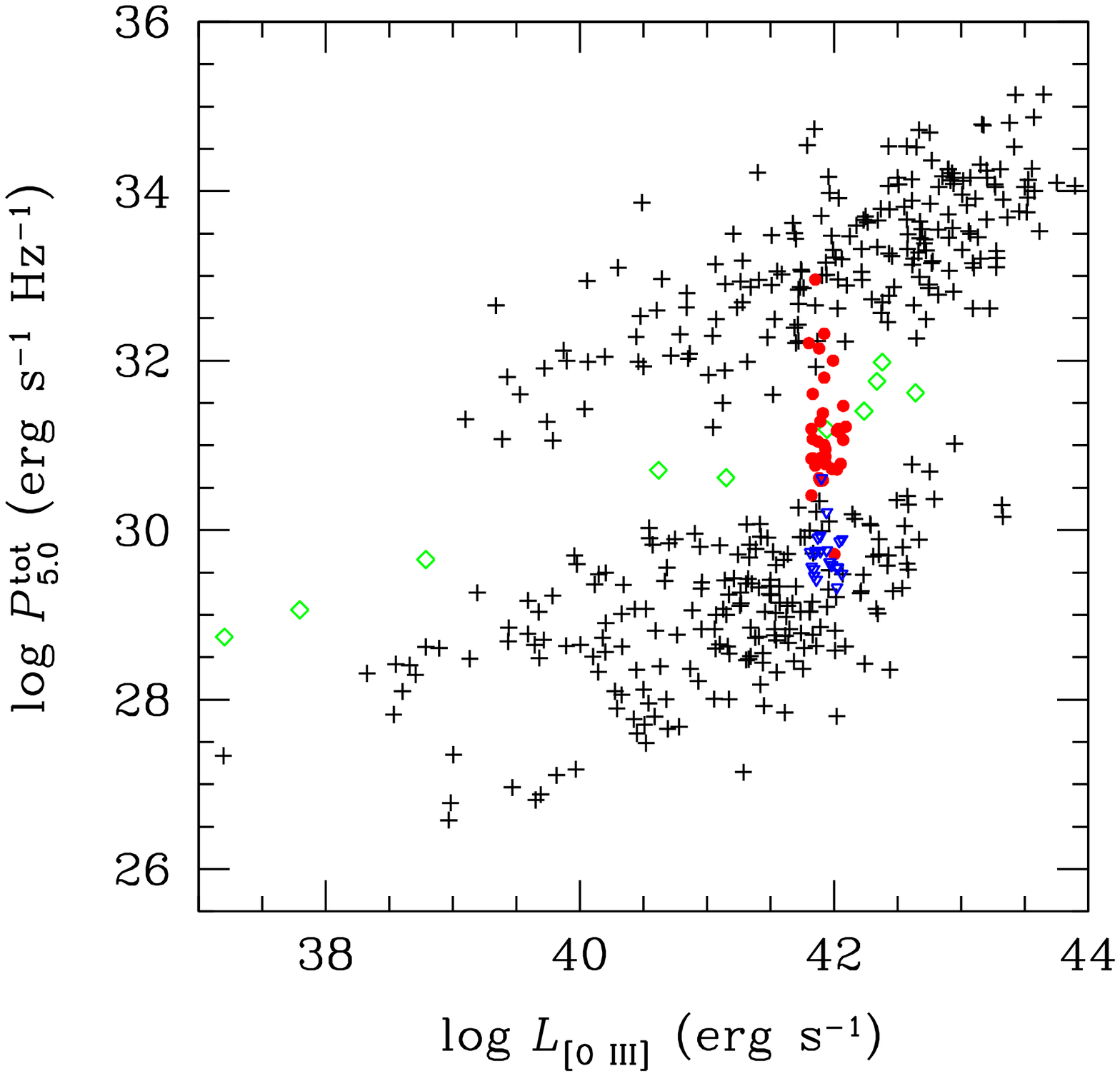}
\caption{Correlation between radio power and \oiii\ luminosity.  The plus 
symbols come from the sample of Xu et~al. (1999), with the open green diamonds 
representing their radio-intermediate sources. The data from Xu et al. were 
taken at 5 GHz and have been corrected to our adopted cosmological parameters 
and to our units for the monochromatic radio power.  Our sample sources are 
shown as filled red circles, with upper limits marked as open blue triangles.  
We converted our measurements to 5 GHz using the observed spectral index for 
each detected source; for the nondetections, we use the median spectral index 
of the detected sources.}
\end{figure}

\clearpage
\figurenum{8}
\begin{figure}
\epsscale{1.00}
\plotone{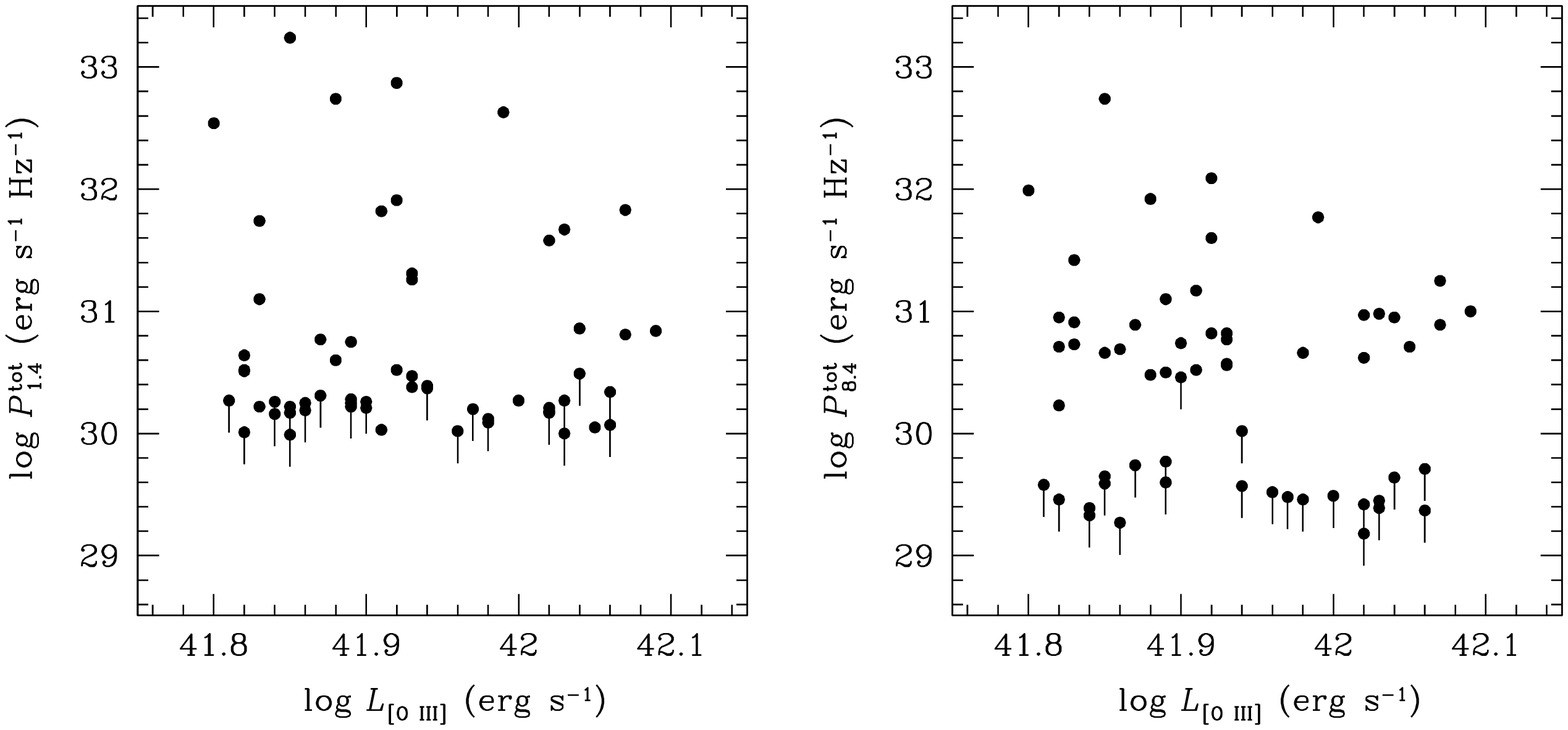}
\caption{Correlation between \oiii\ luminosity versus total integrated radio
power at 1.4 GHz ({\it left}) and 8.4 GHz ({\it right}).  Upper limits are
marked with downward-pointing lines.}
\end{figure}

%TABLES
\clearpage

\begin{deluxetable}{lrl}
\tabletypesize{\footnotesize}
\tablecaption{
\label{log}
Observational Summary
}
\tablewidth{0pt}
\tablehead{
 Parameter &  & Details  %\\
}
\startdata
Observation date  & & 2006 July 24/25 \\
Array configuration  & & B \\
Maximum number of antennas  & & 27 \\
Length of observations (hr)  & & 24 \\
Frequency (GHz)  & & 8.4351 \\
Bandwidth (MHz)  & & 50  \\
Number of IFs  & & 2 \\
Theoretical noise (mJy beam$^{-1}$) &  & 0.02 \\
Primary flux density calibrator  & & 3C\,147/3C\,286 \\
\hspace*{1cm} ... Flux density (Jy) & & 4.74/5.20 \\
Largest angular scale (arcmin)  & & 1 \\
HPBW primary antenna beam (arcmin) & & 5.3 \\
\multicolumn{2}{l}{HPBW synthesized beam} & \\
\hspace*{1cm} ... Uniform weight (arcsec) & & 0.8 \\
\hspace*{1cm} ... Tapered, uniform weight (arcsec) & & 5.4/6.4 \\
\enddata
\end{deluxetable}

\begin{deluxetable}{ccccccccccccc}
\tabletypesize{\scriptsize}
\rotate

\newdimen\digitwidth      % These five lines change the meaning of
\setbox1=\hbox{0}       % " in the same way, so that it leaves as
\digitwidth=\wd1        % much space as a digit. (Note: All digits
\catcode`"=\active      % are the same width).
\def"{\kern\digitwidth}

\tablecaption{
\label{rad_res}
Type 2 Quasar Sample and Radio Measurements
}
\tablewidth{0pt}
\tablehead{
\colhead{Object}                     &      
\multicolumn{2}{c}{Map Parameters}   &     
                                     & 
\multicolumn{9}{c}{Source Parameters} \\
\cline {2-3} \cline{5-13} \\ [-0.2cm]
                                     &              
                                     &                      
                                     &      
                                     &  
\multicolumn{3}{c}{High Resolution}  & 
                                     &  
\multicolumn{5}{c}{Matched Resolution} \\
 \cline{5-7} \cline{9-13} \\ [-0.2cm]
                                     &   
\colhead{Beam}                       & 
\colhead{P.A.}                       &                         
                                     & 
\colhead{\h $S_{8.4}$}               & 
\colhead{\h rms}                     & 
\colhead{\h Notes}                   & 
                                     & 
\colhead{\h $S_{1.4}$}               & 
\colhead{\h rms}                     &
\colhead{\h $S_{8.4}$}               &   
\colhead{\h rms}                     &
\colhead{$\alpha^{8.4}_{1.4}$}       \\
                                     & 
\colhead{(arcsec$^2$)}               &
\colhead{(deg.)}                     & 
                                     &
\colhead{\h (mJy)}                   &
\colhead{\h (mJy~beam$^{-1}$)}       & 
\colhead{\h       }                  & 
                                     & 
\colhead{\h (mJy)}                   & 
\colhead{\h (mJy~beam$^{-1}$)}       & 
\colhead{\h (mJy)}                   & 
\colhead{\h (mJy~beam$^{-1}$)}       & 
                                     \\
(1) & (2) & (3) &&\h  (4) &\h  (5) &\h  (6) &&\h  (7) &\h  (8) &\h  (9) &\h  (10) & (11)  
}
\startdata
SDSS\,J002852.87$-$001433.6 & 0.84$\times$0.69 &151.1 & &\h ""$<$0.110 &\h 0.022 &\h                  & &\h ""$<$0.750 &\h 0.150 &\h " $<$0.210 &\h  0.023 &          \\
SDSS\,J004020.31$-$004033.5 & 1.00$\times$0.72 &144.1 & &\h  ""118.0"" &\h 0.028 &\h   total          & &\h  ""386.6"" &\h 0.165 &\h  ""122.2"" &\h  0.158 & $-$0.641 \\
                            &                  &      & &\h   ""79.54" &\h 0.028 &\h   core+jet       & &\h            &\h       &\h            &\h        &          \\
                            &                  &      & &\h   """6.378 &\h 0.028 &\h   south-east     & &\h            &\h       &\h            &\h        &          \\
                            &                  &      & &\h  """32.30" &\h 0.028 &\h   north-west     & &\h            &\h       &\h            &\h        &          \\
SDSS\,J004412.87$+$003606.8 & 0.92$\times$0.70 &145.6 & &\h ""$<$0.105 &\h 0.021 &\h                  & &\h ""$<$0.750 &\h 0.150 &\h ""$<$0.200 &\h  0.025 &          \\
SDSS\,J013856.14$+$003437.4 & 0.89$\times$0.70 &147.2 & &\h ""$<$0.070 &\h 0.014 &\h                  & &\h ""$<$0.750 &\h 0.150 &\h ""$<$0.180 &\h  0.017 &          \\
SDSS\,J013801.57$-$004946.5 & 0.98$\times$0.72 &146.4 & &\h ""$<$0.170 &\h 0.034 &\h                  & &\h ""$<$0.750 &\h 0.150 &\h ""$<$0.230 &\h  0.036 &          \\
SDSS\,J022004.64$+$005908.3 & 0.78$\times$0.68 &157.9 & &\h ""$<$0.085 &\h 0.017 &\h                  & &\h ""$<$0.750 &\h 0.150 &\h ""$<$0.090 &\h  0.018 &          \\
SDSS\,J022214.12$+$004527.5 & 0.73$\times$0.68 &160.2 & &\h ""$<$0.050 &\h 0.010 &\h                  & &\h ""$<$0.750 &\h 0.150 &\h ""$<$0.140 &\h  0.020 &          \\
SDSS\,J022344.01$+$003914.5 & 0.82$\times$0.66 &177.3 & &\h ""$<$0.100 &\h 0.020 &\h                  & &\h ""$<$0.750 &\h 0.150 &\h ""$<$0.200 &\h  0.022 &          \\
SDSS\,J024240.92$+$004612.1 & 0.86$\times$0.69 &152.8 & &\h   """2.221 &\h 0.020 &\h                  & &\h   """0.943 &\h 0.150 &\h   """2.598 &\h  0.094 & $+$0.564 \\
SDSS\,J024503.71$+$004322.3 & 0.95$\times$0.71 &142.4 & &\h ""$<$0.045 &\h 0.015 &\h                  & &\h ""$<$0.750 &\h 0.150 &\h ""$<$0.210 &\h  0.042 &          \\
SDSS\,J024607.92$-$000532.0 & 0.77$\times$0.68 &167.3 & &\h   """2.002 &\h 0.028 &\h                  & &\h   """0.798 &\h 0.123 &\h   """2.202 &\h  0.077 & $+$0.565 \\
SDSS\,J024919.01$-$000722.5 & 0.75$\times$0.68 &166.7 & &\h ""$<$0.255 &\h 0.051 &\h                  & &\h ""$<$0.750 &\h 0.150 &\h ""$<$0.315 &\h  0.063 &          \\
SDSS\,J025725.99$-$063205.4 & 1.27$\times$0.69 &139.7 & &\h   """3.478 &\h 0.035 &\h   total          & &\h  """77.23" &\h 0.149 &\h  """21.73" &\h  0.296 & $-$0.705 \\
SDSS\,J030809.79$+$005225.8 & 0.74$\times$0.68 &161.0 & &\h ""$<$0.095 &\h 0.019 &\h                  & &\h ""$<$0.750 &\h 0.150 &\h ""$<$0.100 &\h  0.020 &          \\
SDSS\,J031645.60$-$005931.0 & 0.75$\times$0.68 &165.8 & &\h   """0.989 &\h 0.020 &\h                  & &\h   """7.560 &\h 0.164 &\h   """1.518 &\h  0.241 & $-$0.893 \\
SDSS\,J031946.03$-$001629.1 & 0.76$\times$0.69 &161.4 & &\h ""$<$0.105 &\h 0.021 &\h                  & &\h ""$<$0.750 &\h 0.150 &\h ""$<$0.110 &\h  0.022 &          \\
SDSS\,J031947.27$-$010504.0 & 0.78$\times$0.70 &171.1 & &\h ""$<$0.095 &\h 0.019 &\h                  & &\h ""$<$0.750 &\h 0.150 &\h ""$<$0.105 &\h  0.021 &          \\
SDSS\,J032021.94$-$075020.1 & 0.85$\times$0.66 &179.0 & &\h ""$<$0.100 &\h 0.020 &\h                  & &\h ""$<$0.750 &\h 0.150 &\h ""$<$0.101 &\h  0.020 &          \\
SDSS\,J032939.85$+$005220.0 & 0.75$\times$0.69 &154.7 & &\h  """22.27" &\h 0.020 &\h                  & &\h  ""153.7\tablenotemark{a}"" &\h  0.450 &\h  """23.05" &\h  0.665 & $-$1.055 \\
SDSS\,J074130.51$+$302005.3 & 0.68$\times$0.60 &137.0 & &\h   """2.973 &\h 0.012 &\h   total          & &\h   """1.122 &\h 0.156 &\h   """3.060 &\h  0.095 & $+$0.558 \\
SDSS\,J074254.90$+$344236.5 & 0.72$\times$0.60 &123.0 & &\h   """2.040 &\h 0.021 &\h                  & &\h   """1.157 &\h 0.145 &\h   """2.330 &\h  0.082 & $+$0.389 \\
SDSS\,J075238.68$+$390304.9 & 0.70$\times$0.59 &132.1 & &\h   """2.601 &\h 0.013 &\h                  & &\h   """1.894 &\h 0.157 &\h   """2.745 &\h  0.089 & $+$0.206 \\
SDSS\,J081858.36$+$395839.8 & 0.74$\times$0.60 &125.3 & &\h   """3.376 &\h 0.012 &\h   total          & &\h   """5.509 &\h 0.135 &\h   """3.544 &\h  0.108 & $-$0.245 \\
SDSS\,J082449.27$+$370355.7 & 0.75$\times$0.60 &115.5 & &\h   """2.083 &\h 0.021 &\h                  & &\h   """2.817 &\h 0.145 &\h   """2.168 &\h  0.077 & $-$0.146 \\
SDSS\,J083620.35$+$470357.3 & 0.72$\times$0.59 &141.4 & &\h ""$<$0.065 &\h 0.013 &\h                  & &\h ""$<$0.750 &\h 0.150 &\h ""$<$0.070 &\h  0.014 &          \\
SDSS\,J090226.74$+$545952.3 & 0.99$\times$0.58 &101.4 & &\h   """3.380 &\h 0.014 &\h   total          & &\h  """16.89" &\h 0.143 &\h   """3.412 &\h  0.044 & $-$0.890 \\ 
                            &                  &      & &\h ""$<$0.070 &\h 0.014 &\h   south-east     & &\h  """10.25" &\h 0.143 &\h ""$<$0.220 &\h  0.044 &          \\
                            &                  &      & &\h   """3.310 &\h 0.014 &\h   north-west     & &\h   """6.641 &\h 0.143 &\h   """3.392 &\h  0.044 &          \\
SDSS\,J090307.84$+$021152.2 & 0.98$\times$0.72 &137.8 & &\h   """4.603 &\h 0.021 &\h                  & &\h  """20.09" &\h 0.150 &\h   """4.900 &\h  0.151 & $-$0.785 \\
SDSS\,J090801.32$+$434722.6 & 0.75$\times$0.62 &"57.7 & &\h   """6.229 &\h 0.023 &\h                  & &\h  """28.95" &\h 0.150 &\h   """6.408 &\h  0.176 & $-$0.839 \\
SDSS\,J094209.00$+$570019.7 & 0.87$\times$0.57 &118.4 & &\h   """2.284 &\h 0.022 &\h                  & &\h   """0.775 &\h 0.145 &\h   """2.350 &\h  0.083 & $+$0.617 \\
SDSS\,J094350.92$+$610255.9 & 0.86$\times$0.57 &126.3 & &\h ""$<$0.135 &\h 0.027 &\h                  & &\h ""$<$0.750 &\h 0.150 &\h ""$<$0.150 &\h  0.030 &          \\
SDSS\,J094557.03$+$570803.2 & 0.81$\times$0.57 &131.2 & &\h  """10.99" &\h 0.013 &\h   total          & &\h  """23.48" &\h 0.146 &\h  """11.58" &\h  0.313 & $-$0.393 \\
SDSS\,J095044.69$+$011127.2 & 1.54$\times$0.72 &129.2 & &\h   """2.408 &\h 0.024 &\h                  & &\h   """1.719 &\h 0.128 &\h   """2.700 &\h  0.102 & $+$0.251 \\
SDSS\,J095629.06$+$573508.9 & 0.92$\times$0.57 &111.7 & &\h   """2.877 &\h 0.014 &\h   total          & &\h   """0.880 &\h 0.150 &\h   """3.218 &\h  0.100 & $+$0.721 \\
SDSS\,J095941.73$+$580545.9 & 0.77$\times$0.57 &148.4 & &\h ""$<$0.135 &\h 0.027 &\h                  & &\h ""$<$0.750 &\h 0.150 &\h ""$<$0.155 &\h  0.031 &          \\
SDSS\,J100854.43$+$461300.7 & 1.16$\times$0.63 &"97.2 & &\h  """21.68" &\h 0.018 &\h   total          & &\h  ""149.3"" &\h 0.147 &\h  """25.22" &\h  0.053 & $-$0.989 \\
                            &                  &      & &\h   """0.352 &\h 0.018 &\h   core+jet       & &\h   """4.344 &\h 0.147 &\h   """0.492 &\h  0.053 &          \\
                            &                  &      & &\h   """2.726 &\h 0.018 &\h   north-east     & &\h  """30.64" &\h 0.147 &\h   """4.824 &\h  0.053 &          \\
                            &                  &      & &\h  """18.60" &\h 0.018 &\h   south-west     & &\h  ""112.3"" &\h 0.147 &\h  """19.89" &\h  0.053 &          \\
SDSS\,J101237.32$+$023554.3 & 1.05$\times$0.72 &133.7 & &\h   """2.371 &\h 0.012 &\h                  & &\h   """0.957 &\h 0.150 &\h   """2.683 &\h  0.098 & $+$0.573 \\
SDSS\,J101403.49$+$024416.4 & 1.33$\times$0.71 &"53.8 & &\h   """4.532 &\h 0.023 &\h   total          & &\h   """2.023 &\h 0.126 &\h   """4.537 &\h  0.168 & $+$0.449 \\
SDSS\,J101718.63$+$033108.2 & 0.96$\times$0.71 &135.7 & &\h   """2.847 &\h 0.021 &\h                  & &\h   """2.543 &\h 0.150 &\h   """3.348 &\h  0.109 & $+$0.153 \\
SDSS\,J102640.42$-$004206.5 & 1.64$\times$0.72 &126.1 & &\h   """3.677 &\h 0.019 &\h   total          & &\h   """1.852 &\h 0.150 &\h   """3.679 &\h  0.125 & $+$0.380 \\
SDSS\,J103639.39$+$640924.7 & 0.80$\times$0.56 &154.2 & &\h ""$<$0.130 &\h 0.026 &\h                  & &\h ""$<$0.750 &\h 0.150 &\h ""$<$0.135 &\h  0.027 &          \\
SDSS\,J111112.87$+$030850.3 & 0.92$\times$0.77 &"13.6 & &\h ""$<$0.555 &\h 0.111 &\h                  & &\h ""$<$0.750 &\h 0.150 &\h ""$<$1.175 &\h  0.235 &          \\
SDSS\,J121856.42$+$611922.7 & 1.04$\times$0.60 &"81.3 & &\h ""$<$0.155 &\h 0.031 &\h                  & &\h ""$<$0.750 &\h 0.150 &\h ""$<$0.165 &\h  0.033 &          \\
SDSS\,J124749.79$+$015212.6 & 1.98$\times$0.69 &126.7 & &\h   """9.203 &\h 0.022 &\h   total          & &\h  """21.02" &\h 0.143 &\h   """10.11"&\h  0.066 & $-$0.407 \\ 
                            &                  &      & &\h   """8.983 &\h 0.022 &\h   core           & &\h   """6.487 &\h 0.143 &\h   """9.453 &\h  0.066 &          \\
                            &                  &      & &\h ""$<$0.110 &\h 0.022 &\h   east           & &\h   """7.333 &\h 0.143 &\h ""$<$0.330 &\h  0.066 &          \\
                            &                  &      & &\h ""$<$0.110 &\h 0.022 &\h   north-west     & &\h   """7.201 &\h 0.143 &\h ""$<$0.330 &\h  0.066 &          \\
SDSS\,J135128.14$-$001016.9 & 0.79$\times$0.71 &"17.2 & &\h   """2.231 &\h 0.021 &\h                  & &\h   """2.223 &\h 0.149 &\h   """2.630 &\h  0.091 & $+$0.094 \\
SDSS\,J143027.66$-$005614.9 & 0.78$\times$0.70 &""7.4 & &\h ""$<$0.230 &\h 0.046 &\h                  & &\h ""$<$0.750 &\h 0.150 &\h ""$<$0.235 &\h  0.047 &          \\
SDSS\,J143047.33$+$602304.5 & 1.03$\times$0.57 &100.1 & &\h   """2.551 &\h 0.022 &\h                  & &\h   """2.080 &\h 0.150 &\h   """2.549 &\h  0.095 & $+$0.113 \\
SDSS\,J144711.29$+$021136.2 & 0.73$\times$0.68 &151.3 & &\h   """2.757 &\h 0.012 &\h   total          & &\h   """0.762 &\h 0.150 &\h   """3.487 &\h  0.114 & $+$0.846 \\
SDSS\,J152019.75$-$013611.2 & 0.77$\times$0.71 &165.9 & &\h   """2.128 &\h 0.019 &\h                  & &\h   """1.411 &\h 0.146 &\h   """2.551 &\h  0.086 & $+$0.329 \\
SDSS\,J153734.00$+$511258.9 & 0.74$\times$0.60 &136.5 & &\h   """2.646 &\h 0.022 &\h                  & &\h   """0.861 &\h 0.138 &\h   """2.962 &\h  0.100 & $+$0.687 \\
SDSS\,J153943.73$+$514221.0 & 0.76$\times$0.61 &131.4 & &\h   """2.817 &\h 0.023 &\h                  & &\h  """13.57" &\h 0.148 &\h   """3.566 &\h  0.139 & $-$0.743 \\
SDSS\,J154133.19$+$521200.1 & 0.80$\times$0.59 &125.1 & &\h ""$<$0.330 &\h 0.066 &\h                  & &\h ""$<$0.750 &\h 0.150 &\h ""$<$0.340 &\h  0.068 &          \\
SDSS\,J154337.82$-$004419.9 & 0.79$\times$0.68 &166.4 & &\h ""$<$0.155 &\h 0.031 &\h                  & &\h   """1.017 &\h 0.137 &\h ""$<$0.170 &\h  0.034 & $< -$0.995 \\
SDSS\,J154826.05$+$004615.3 & 0.75$\times$0.69 &164.6 & &\h   """0.748 &\h 0.019 &\h                  & &\h   """4.208 &\h 0.153 &\h   """0.760 &\h  0.134 & $-$0.954 \\
SDSS\,J173938.64$+$544208.6 & 0.79$\times$0.56 &140.6 & &\h   """2.558 &\h 0.022 &\h                  & &\h   """0.948 &\h 0.151 &\h   """2.594 &\h  0.098 & $+$0.560 \\
SDSS\,J215731.40$+$003757.1 & 0.84$\times$0.68 &158.7 & &\h  """20.18" &\h 0.018 &\h   total          & &\h  ""148.86" &\h 0.146 &\h  """20.56" &\h  0.128 & $-$1.101 \\
                            &                  &      & &\h  """13.96" &\h 0.018 &\h   east           & &\h  """89.03" &\h 0.146 &\h  """14.25" &\h  0.128 &          \\
                            &                  &      & &\h   """3.198 &\h 0.018 &\h   west           & &\h  """45.96" &\h 0.146 &\h   """3.292 &\h  0.128 &          \\
                            &                  &      & &\h   """3.023 &\h 0.018 &\h   south          & &\h  """13.88" &\h 0.146 &\h   """3.024 &\h  0.128 &          \\
SDSS\,J222631.14$-$010054.0 & 1.00$\times$0.67 &168.7 & &\h ""$<$0.170 &\h 0.034 &\h                  & &\h ""$<$0.750 &\h 0.150 &\h ""$<$0.175 &\h  0.035 &          \\
SDSS\,J225227.39$-$005528.5 & 0.90$\times$0.68 &155.7 & &\h   """2.318 &\h 0.020 &\h                  & &\h   """0.892 &\h 0.127 &\h   """2.880 &\h  0.103 & $+$0.652 \\
SDSS\,J230937.14$+$001735.8 & 0.89$\times$0.68 &156.0 & &\h ""$<$0.115 &\h 0.023 &\h                  & &\h ""$<$0.750 &\h 0.150 &\h ""$<$0.120 &\h  0.024 &          \\
SDSS\,J235433.86$-$005629.3 & 0.94$\times$0.71 &149.5 & &\h   """0.694 &\h 0.027 &\h                  & &\h   """2.251 &\h 0.142 &\h   """0.877 &\h  0.071 & $-$0.524 \\
\enddata
\tablenotetext{a}{Not included in FIRST; the listed 1.4 GHz data come from NVSS.}
\end{deluxetable}

\begin{deluxetable}{cccrrcccc}
\tabletypesize{\footnotesize}
\tablecaption{
\label{summary}
Summary of Radio and Optical Properties
}
\tablewidth{0pt}
\tablehead{
                     &       &        & \multicolumn{3}{c}{Radio Properties}   &     & \multicolumn{2}{c}{Optical Properties} \\
 \cline{4-6} \cline{8-9} \\ [-0.2cm]
   Galaxy &  $z$  & $D_L$ &\colhead{log $P^{\rm tot}_{1.4}$} & \colhead{log $P^{\rm tot}_{8.4}$} & Morphology & &  \colhead{$\log L_{\rm [O\,\sc II]}$} & \colhead{$\log L_{\rm [O\,\sc III]}$} \\
                     &       & (Mpc)  & \multicolumn{2}{c}{(erg~s$^{-1}$~Hz$^{-1}$)} & & &  \multicolumn{2}{c}{(erg~s$^{-1}$)} \\
               (1)   & (2)   &  (3)   &  (4)   &  (5) &(6)& & (7)&  (8) 
}
\startdata
SDSS\,J0028$-$0014 & 0.310 & 1231   & $<$30.00 & $<$29.45 &  \nd  & & 41.56 & 42.03  \\
SDSS\,J0040$-$0040 & 0.568 & 2107   &    33.24 &    32.74 &  E    & & 41.61 & 41.85  \\
SDSS\,J0044$+$0036 & 0.502 & 1895   & $<$30.31 & $<$29.74 &  \nd  & & 41.57 & 41.87  \\
SDSS\,J0138$-$0049 & 0.433 & 1665   & $<$30.22 & $<$29.60 &  \nd  & & 41.40 & 41.89  \\
SDSS\,J0138$+$0034 & 0.478 & 1816   & $<$30.28 & $<$29.77 &  \nd  & & 41.62 & 41.89  \\
SDSS\,J0220$+$0059 & 0.413 & 1597   & $<$30.19 & $<$29.27 &  \nd  & & 41.70 & 41.86  \\
SDSS\,J0222$+$0045 & 0.421 & 1624   & $<$30.20 & $<$29.48 &  \nd  & & 41.44 & 41.97  \\
SDSS\,J0223$+$0039 & 0.397 & 1541   & $<$30.17 & $<$29.59 &  \nd  & & 41.48 & 41.85  \\
SDSS\,J0242$+$0046 & 0.408 & 1579   &    30.22 &    30.66 &  U    & & 41.86 & 41.85  \\
SDSS\,J0245$+$0043 & 0.315 & 1249   & $<$30.01 & $<$29.46 &  \nd  & & 41.23 & 41.82  \\
SDSS\,J0246$-$0005 & 0.493 & 1867   &    30.25 &    30.69 &  U    & & 42.05 & 41.86  \\
SDSS\,J0249$-$0007 & 0.579 & 2141   & $<$30.39 & $<$30.02 &  \nd  & & 41.70 & 41.94  \\
SDSS\,J0257$-$0632 & 0.557 & 2072   &    32.54 &    31.99 &  L    & & 41.40 & 41.80  \\
SDSS\,J0308$+$0052 & 0.466 & 1776   & $<$30.27 & $<$29.39 &  \nd  & & 41.54 & 42.03  \\
SDSS\,J0316$-$0059 & 0.369 & 1443   &    31.26 &    30.56 &  U    & & 41.63 & 41.93  \\
SDSS\,J0319$-$0016 & 0.393 & 1527   & $<$30.16 & $<$29.33 &  \nd  & & 41.17 & 41.84  \\
SDSS\,J0319$-$0105 & 0.699 & 2503   & $<$30.49 & $<$29.64 &  \nd  & & 41.80 & 42.04  \\
SDSS\,J0320$-$0750 & 0.457 & 1746   & $<$30.26 & $<$29.39 &  \nd  & & 41.49 & 41.84  \\
SDSS\,J0329$+$0052 & 0.446 & 1709   &    32.74 &    31.92 &  S    & & 42.14 & 41.88  \\
SDSS\,J0741$+$3020 & 0.476 & 1810   &    30.38 &    30.82 &  S    & & 41.71 & 41.93  \\
SDSS\,J0742$+$3442 & 0.567 & 2103   &    30.52 &    30.82 &  U    & & 41.48 & 41.92  \\
SDSS\,J0752$+$3903 & 0.654 & 2370   &    30.84 &    31.00 &  S    & & 41.32 & 42.09  \\
SDSS\,J0818$+$3958 & 0.406 & 1572   &    31.10 &    30.91 &  S    & & 41.49 & 41.83  \\
SDSS\,J0824$+$3703 & 0.305 & 1213   &    30.60 &    30.48 &  U    & & 41.35 & 41.88  \\
SDSS\,J0836$+$4703 & 0.423 & 1631   & $<$30.21 & $<$29.18 &  \nd  & & 41.13 & 42.02  \\
SDSS\,J0902$+$5459 & 0.401 & 1555   &    31.67 &    30.98 &  L$+$D& & 41.27 & 42.03  \\ 
SDSS\,J0903$+$0211 & 0.329 & 1300   &    31.58 &    30.97 &  S    & & 42.01 & 42.02  \\
SDSS\,J0908$+$4347 & 0.363 & 1422   &    31.82 &    31.17 &  U    & & 41.53 & 41.91  \\
SDSS\,J0942$+$5700 & 0.350 & 1376   &    30.03 &    30.52 &  S    & & 41.92 & 41.91  \\
SDSS\,J0943$+$6102 & 0.341 & 1343   & $<$30.07 & $<$29.37 &  \nd  & & 41.39 & 42.06  \\
SDSS\,J0945$+$5708 & 0.512 & 1928   &    31.91 &    31.60 &  S    & & 41.68 & 41.92  \\
SDSS\,J0950$+$0111 & 0.404 & 1565   &    30.52 &    30.71 &  U    & & 41.65 & 41.82  \\
SDSS\,J0956$+$5735 & 0.361 & 1415   &    30.09 &    30.66 &  S    & & 41.35 & 41.98  \\
SDSS\,J0959$+$5805 & 0.465 & 1773   & $<$30.27 & $<$29.58 &  \nd  & & 41.61 & 41.81  \\
SDSS\,J1008$+$4613 & 0.544 & 2031   &    32.87 &    32.09 &  U$+$L& & 41.44 & 41.92  \\
SDSS\,J1012$+$0235 & 0.720 & 2563   &    30.51 &    30.95 &  U    & & 41.77 & 41.82  \\
SDSS\,J1014$+$0244 & 0.571 & 2116   &    30.75 &    31.10 &  S$+$L& & 41.65 & 41.89  \\
SDSS\,J1017$+$0331 & 0.453 & 1733   &    30.77 &    30.89 &  U    & & 41.26 & 41.87  \\
SDSS\,J1026$-$0042 & 0.365 & 1429   &    30.47 &    30.77 &  S    & & 41.65 & 41.93  \\
SDSS\,J1036$+$6409 & 0.398 & 1545   & $<$30.17 & $<$29.42 &  \nd  & & 41.48 & 42.02  \\
SDSS\,J1111$+$0308 & 0.461 & 1760   & $<$30.26 & $<$30.46 &  \nd  & & 41.49 & 41.90  \\
SDSS\,J1218$+$6119 & 0.369 & 1443   & $<$30.12 & $<$29.46 &  \nd  & & 41.65 & 41.98  \\
SDSS\,J1247$+$0152 & 0.427 & 1645   &    31.74 &    31.42 &  E    & & 41.78 & 41.83  \\  
SDSS\,J1351$-$0010 & 0.524 & 1967   &    30.81 &    30.89 &  S    & & 42.09 & 42.07  \\
SDSS\,J1430$-$0056 & 0.318 & 1260   & $<$30.02 & $<$29.52 &  \nd  & & 41.45 & 41.96  \\
SDSS\,J1430$+$6023 & 0.607 & 2228   &    30.86 &    30.95 &  S    & & 41.90 & 42.04  \\
SDSS\,J1447$+$0211 & 0.386 & 1503   &    30.05 &    30.71 &  S    & & 41.70 & 42.05  \\
SDSS\,J1520$-$0136 & 0.307 & 1220   &    30.25 &    30.50 &  U    & & 41.32 & 41.89  \\
SDSS\,J1537$+$5112 & 0.444 & 1702   &    30.21 &    30.74 &  U    & & 41.13 & 41.90  \\
SDSS\,J1539$+$5142 & 0.585 & 2160   &    31.83 &    31.25 &  S    & & 41.95 & 42.07  \\
SDSS\,J1541$+$5212 & 0.305 & 1213   & $<$29.99 & $<$29.65 &  \nd  & & 41.68 & 41.85  \\
SDSS\,J1543$-$0044 & 0.311 & 1235   &    30.27 & $<$29.49 &  \nd  & & 41.34 & 42.00  \\
SDSS\,J1548$+$0046 & 0.544 & 2031   &    31.31 &    30.57 &  S$+$L& & 41.47 & 41.93  \\
SDSS\,J1739$+$5442 & 0.384 & 1496   &    30.18 &    30.62 &  U    & & 41.16 & 42.02  \\
SDSS\,J2157$+$0037 & 0.390 & 1517   &    32.63 &    31.77 &  E    & & 41.53 & 41.99  \\
SDSS\,J2226$-$0100 & 0.530 & 1986   & $<$30.34 & $<$29.71 &  \nd  & & 41.55 & 42.06  \\
SDSS\,J2252$-$0055 & 0.442 & 1696   &    30.22 &    30.73 &  U    & & 41.62 & 41.83  \\
SDSS\,J2309$+$0017 & 0.555 & 2066   & $<$30.37 & $<$29.57 &  \nd  & & 41.71 & 41.94  \\
SDSS\,J2354$-$0056 & 0.348 & 1368   &    30.64 &    30.23 &  U    & & 41.37 & 41.82  \\
\enddata
\end{deluxetable}

\end{document}